\documentclass[twocolumn]{aastex6}
\usepackage{graphicx}
\usepackage{amssymb,amsfonts,amsmath,amstext,amsgen,amsopn,amsxtra,indentfirst,times,longtable}

\begin{document}


\title{The Gradient of Mean Molecular Weight Across the Radius Valley}

\author{Kevin Heng\altaffilmark{1,2,3,4}}
\author{James E. Owen\altaffilmark{5}}
\author{Meng Tian\altaffilmark{1}}
\altaffiltext{1}{Faculty of Physics, Ludwig Maximilian University, Scheinerstrasse 1, D-81679, Munich, Bavaria, Germany. \\ Emails: Kevin.Heng@physik.lmu.de, Meng.Tian@physik.lmu.de}
\altaffiltext{2}{ARTORG Center for Biomedical Engineering Research, University of Bern, Murtenstrasse 50, CH-3008, Bern, Switzerland}
\altaffiltext{3}{University College London, Department of Physics \& Astronomy, Gower St, London, WC1E 6BT, United Kingdom}
\altaffiltext{4}{Astronomy \& Astrophysics Group, Department of Physics, University of Warwick, Coventry CV4 7AL, United Kingdom}
\altaffiltext{5}{Astrophysics Group, Department of Physics, Imperial College London, Prince Consort Rd, London SW7 2AZ, United Kingdom}

\begin{abstract}
Photo-evaporation shapes the observed radii of small exoplanets and constrains the underlying distributions of atmospheric and core masses.  However, the diversity of atmospheric chemistries corresponding to these distributions remains unelucidated.  We develop a first-principles carbon-hydrogen-oxygen-sulfur-silicon (CHOSSi) outgassing model that accounts for non-ideal gas behavior (via fugacities) at high pressures, as well as the tendency for water and hydrogen to dissolve in melt (via solubility laws).  We use data-driven radius valley constraints to establish the relationship between the atmospheric surface pressures and melt temperatures of sub-Neptunes.  Sub-Neptunes with less massive rocky cores retain less of their primordial hydrogen envelopes, which leads to less heat retention and diminished melt temperatures at the surfaces of these cores.  Lower melt temperatures lead thermodynamically to the dominance of carbon-, oxygen-, sulfur- and silicon-bearing molecules over molecular hydrogen, which naturally produce a diversity of mean molecular weights.  Our geochemical outgassing calculations robustly predict a gradient of mean molecular weight across the radius valley, where the strength of this gradient is primarily driven by the oxygen fugacity of the molten cores and not by the carbon enrichment (or ``metallicity") of the atmosphere.  Smaller sub-Neptunes are predicted to have less hydrogen-dominated atmospheres.  Establishing the precise relationship between the observed and outgassed chemistries requires an understanding of how convection near the core interacts with large-scale atmospheric circulation (driven by stellar heating) near the photosphere, as well as the influence of photochemistry.
\end{abstract}


\section{Introduction}
\label{sect:intro}

Exoplanets intermediate in size between Earth and Neptune are common \citep{fulton17}.  They cluster mainly into two categories separated by a ``radius valley" \citep{fp18,lp22}: objects with bulk densities high enough to be dominated by a rock-metal core (super Earths) versus those with a voluminous hydrogen-helium envelope of probably primordial origin (sub-Neptunes).  It is likely that super Earths and sub-Neptunes originate from the same underlying population of exoplanets with the former losing their primordial atmospheres via photo-evaporation \citep{ow13,ow16} and/or core-powered mass loss \citep{ginzburg18,gs19}.

 In the Solar System, the terrestrial planets (and moons) have secondary atmospheres sourced by geochemical outgassing (e.g., \citealt{gs14,gaillard22}), while the gas and ice giants have primary atmospheres that consist of primordial hydrogen and helium from the protostellar disk.  The hydrogen and helium content of the atmospheres of sub-Neptunes span a continuous range of masses and thus surface pressures, due to the varying extent to which they have retained these atmospheres against X-ray and extreme ultraviolet (EUV) driven atmospheric escape \citep{ro21}.  This implies that some of them have hybrid atmospheres, where geochemical outgassing occurs in the presence of primordial hydrogen and helium \citep{th24}.  
 
 \begin{figure}[!ht]
\begin{center}
\vspace{-0.1in} 
\includegraphics[width=\columnwidth]{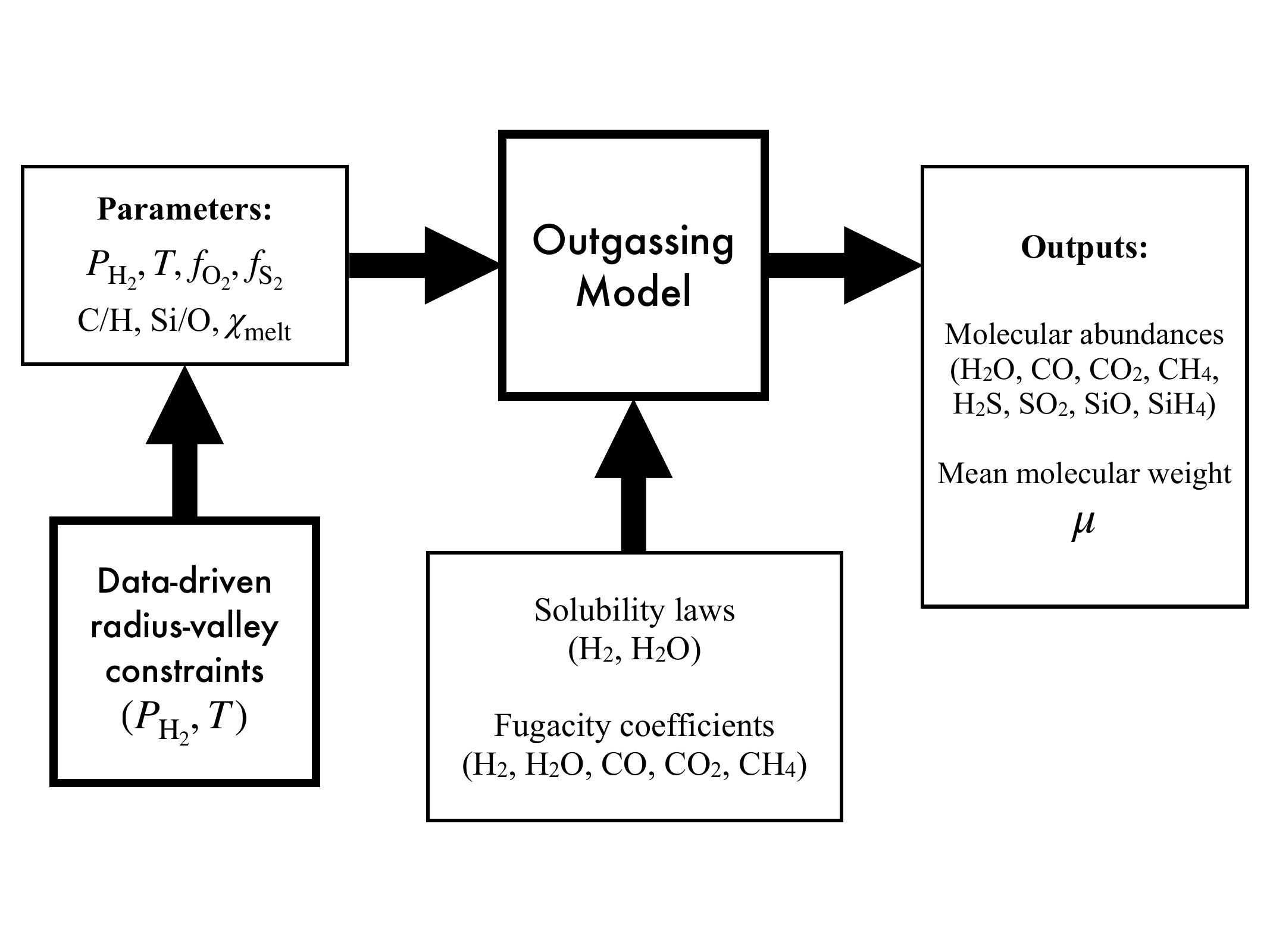}
\end{center}
\vspace{-0.3in}
\caption{Schematic describing how geochemical outgassing calculations are combined with data-driven radius valley constraints to produce molecular abundances and the mean molecular weight.}
\label{fig:schematic}
\end{figure}
 
Previously, \cite{ro21} derived the distributions of atmospheric mass fractions and core masses in the population of super Earths and sub-Neptunes based on matching the observed radius versus orbital period distributions.  The atmospheric mass fractions may be converted into atmospheric surface pressures.  With these atmospheric surface pressures in hand, one may use improved geochemical outgassing models \citep{th24} to predict the plausible range of atmospheric chemistries.  In the current study, we pursue exactly this approach (Figure \ref{fig:schematic}) focusing on sub-Neptunes above the radius valley (with atmospheric pressures $\gtrsim 1000$ bar).  From the predicted range of atmospheric chemistries, we wish to understand the distribution of mean molecular weights predicted,
\begin{equation}
\mu = 2X_{\rm H_2} + 4 X_{\rm He} + \sum_i X_i \mu_i,
\end{equation}
where the summation is over the molecular weight ($\mu_i$) and volume mixing ratios or mole fractions ($X_i$) of the other atmospheric atoms and molecules.  Primordial hydrogen-helium atmospheres have $\mu \approx 2.2$ ($X_{\rm H_2} = 0.9$, $X_{\rm He} = 0.1 X_{\rm H_2}$), while the Earth's nitrogen-dominated ($X_{\rm N_2} = 0.78$, $X_{\rm O_2} = 0.21$) secondary atmosphere has $\mu \approx 29$.  The mean molecular weight of the atmosphere of Jupiter\footnote{\texttt{https://nssdc.gsfc.nasa.gov/planetary/factsheet\\/jupiterfact.html}} is about 2.2, while it is slightly lower (about 2.1) for Saturn\footnote{\texttt{https://nssdc.gsfc.nasa.gov/planetary/factsheet\\/saturnfact.html}} due to the relative scarcity of helium.

With mean molecular weights of intermediate values being reported for small exoplanets (e.g., $\mu \approx 5$--$6$ for the sub-Neptune TOI-270d; \citealt{benneke24,felix25}), it becomes relevant to understand if hybrid atmospheres may account for them.  As the primordial hydrogen- helium envelope becomes more massive, we expect the mean molecular weight to tend towards $\mu \approx 2.2$.

Key questions we wish to address include:
\begin{itemize}

    \item What is the threshold radius beyond which the mean molecular weight starts to depart from the canonical value for hydrogen-dominated atmospheres?

    \item Is there a \textit{gradient} of mean molecular weight across planetary radius and orbital period?

    \item What is the physical mechanism that establishes this gradient of mean molecular weight?

    \item What is the key parameter that drives the \textit{strength} of this gradient of mean molecular weight?

    \item Can methane and carbon dioxide be generated in comparable amounts by outgassing?
    
\end{itemize}

In Section \ref{sect:methods}, we describe our methodology including how we establish the temperature-pressure conditions at the surfaces of sub-Neptunian cores and an improved CHOSSi outgassing model that accounts for fugacities and solubilities.  In Section \ref{sect:results}, we present our calculations of outgassing, first by themselves and later coupled to data-driven radius valley constraints.  In Section \ref{sect:discussion}, we discuss the implications of our results and suggest opportunities for future work.

\section{Methodology}
\label{sect:methods}

\subsection{Inferred properties of small exoplanet population}

\cite{ro21} generated a synthetic population of exoplanets and subjected it to 3 Gyr of X-ray and EUV driven photo-evaporation.  By matching this synthetic population to the observed radius versus orbital period distribution of super Earths and sub-Neptunes \citep{fp18}, they were able to constrain the distributions of core masses, core densities and final atmospheric mass fractions.  This data-driven inference approach revealed a distribution of core masses peaking at about 4 Earth masses and a distribution of atmospheric mass fractions peaking at about 2\%.   The \cite{ro21} models used interior structure evolution calculations which self-consistently generated the surface pressures and temperatures. Although these values were not reported directly in their work, we make use of these core-atmosphere interface temperatures to inform our outgassing calculations.  We use the surface pressure to inform the partial pressure of molecular hydrogen as \cite{ro21} assumed hydrogen-dominated atmospheres in their calculations.

\subsection{Carbon-hydrogen-oxygen-sulfur-silicon (CHOSSi) outgassing model}

\cite{th24} previously demonstrated that the same outgassing model may be used to compute the atmospheric chemistry of both secondary and hybrid atmospheres, where one assumes the total surface pressure and hydrogen partial pressure, respectively.  Following \cite{french66}, they parametrized a thermodynamic activity associated with graphite to describe the carbon content of the melt.  Their framework and calculations considered a carbon-hydrogen-oxygen-nitrogen-sulfur (CHONS) chemical system and accounted for non-ideal gases (via fugacity coefficients) and non-ideal mixing of gaseous components (via activity coefficients).  However, they did not include the effect of the gases dissolving into the melt (as expressed through solubility laws).  In the current study, we account for the solubility of water and molecular hydrogen, which requires the mass budget of hydrogen to be explicitly considered.

\subsubsection{Solubility laws}
\label{subsect:solubility}

Outgassing models typically describe a system consisting of a mixed gas (the atmosphere) and a liquid (the melt) in collective chemical equilibrium (e.g., \citealt{french66,gs14,gaillard22,th24}).  Some of these gases are partially dissolved in the melt, a phenomenon that is well established in the geosciences (e.g., \citealt{gaillard22}).  In the current study, we take the approximation that only water (H$_2$O) and molecular hydrogen (H$_2$) have non-negligible solubilities.  The former is well established in the geosciences literature (e.g., \citealt{mcmillan94,berndt02,dufils20}).  The latter is motivated by the claim that objects with sizes larger than about three Earth radii are rare because of the dissolution of molecular hydrogen in melt under high pressure---the so-called ``fugacity crisis" \citep{kite19}.  Under Earth-like conditions, carbon dioxide (CO$_2$), carbon monoxide (CO) and methane (CH$_4$) have lower solubilities, compared to water, in basaltic melts (e.g., \citealt{amalberti21,seo24}).  Section 2.3 of \cite{seo24} contains a comprehensive discussion of the solubility laws associated with CO$_2$, CO and CH$_4$.  Using scaling arguments, \cite{seo24} conclude that the dissolution of CO and CH$_4$ into the melt is negligible, while CO$_2$ is moderately soluble when the atmosphere is ``highly oxidized" and the silicate layer of the core is ``almost fully molten and well-mixed".  We also ignore the solubility of sulfur species as the solubility laws are unavailable.

\begin{figure}[!ht]
\begin{center}
\vspace{-0.2in} 
\includegraphics[width=\columnwidth]{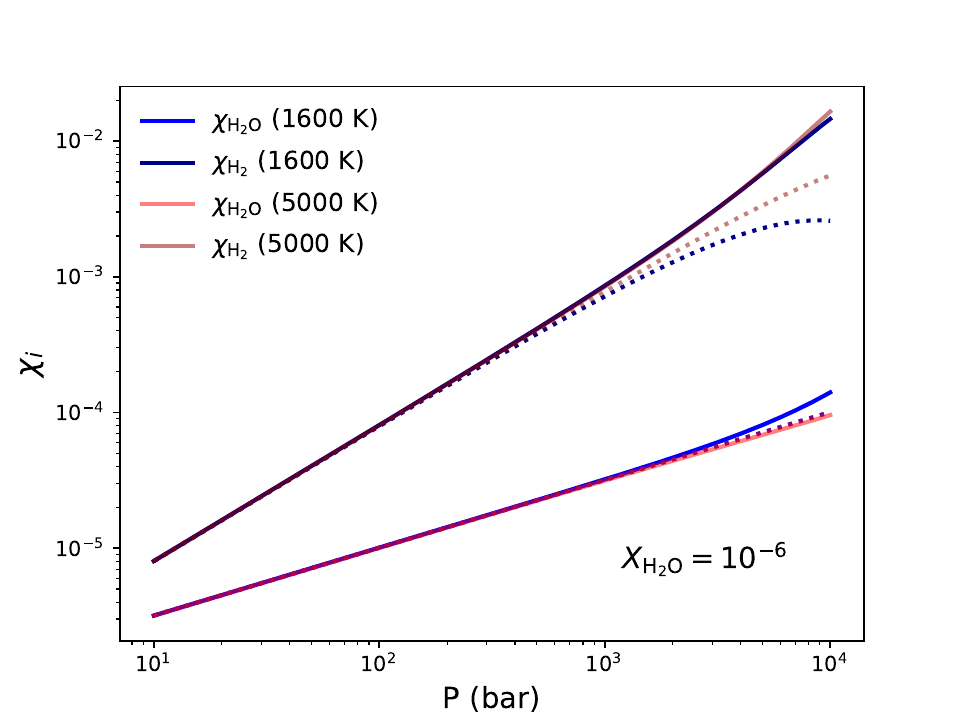}
\end{center}
\caption{Mass fraction of water and hydrogen dissolved in melt as a function of the total ambient pressure.  Shown are temperatures representative of Earth-like melt (1600 K) and sub-Neptune core surfaces (5000 K).  The dotted curves show the same solubility laws but with the fugacity coefficient set to unity ($\phi_i=1$).  For illustration, the volume mixing ratio of water has been set to $X_{\rm H_2O}=10^{-6}$ (for converting $P_{\rm H_2O} = X_{\rm H_2O} P$) and we have assumed $P_{\rm H_2}=P$ for hydrogen.  Sub-Neptunes have characteristic surface pressures of $P\sim 10^4$ bar.}
\label{fig:solubility}
\end{figure}

In the simplest carbon-hydrogen-oxygen (CHO) chemical system, the set of coupled equations reduces to a quadratic equation for the partial pressure of molecular hydrogen \citep{french66}.  When solubility laws for water and molecular hydrogen are considered, this governing equation generalizes to a polynomial equation.  The exact form of this equation depends on the functional form of the solubility laws, which is why it is relevant to establish them prior to deriving the outgassing model.  

For water, we use the solubility law for basaltic melts as compiled in Table 1 of \cite{bower22},
\begin{equation}
\chi_{\rm H_2O} = A_{\rm H_2O} \left( \frac{f_{\rm H_2O}}{P_0} \right)^{1/2},
\end{equation}
where $\chi_{\rm H_2O}$ is the mass fraction of water dissolved in the melt, $f_{\rm H_2O} = \phi_{\rm H_2O} P_{\rm H_2O}$ is the fugacity of water, $\phi_{\rm H_2O}$ is the fugacity coefficient of water, $P_{\rm H_2O}$ is the partial pressure of water in the atmosphere and $P_0 = 1$ bar is the commonly used reference pressure.  The fugacity is a generalization of the partial pressure under non-ideal-gas conditions.  The coefficient of $A_{\rm H_2O} = 1.007 \times 10^{-3}$, as well as the $\chi_{\rm H_2O} \propto P_{\rm H_2O}^{1/2}$ functional dependence, was calibrated at a temperature of 1473 K, a pressure range of 503--2021 bar and a range of oxygen fugacities between IW+3.5 and IW+7.9 \citep{berndt02}.  Strictly speaking, this calibration is not always consistent with the range of pressures and oxygen fugacities explored in the current study.

For molecular hydrogen, we use the solubility law for basaltic melts from \cite{gaillard22}\footnote{See equation (8) of the Supplementary Methods section.  We note a typographical error in the coefficients that switched ``9.43" and ``1.51".  With this correction, the numerical coefficient multiplying $f_{\rm H_2}$ now matches the fitting function stated in equation (3) of \cite{kite19} at the order-of-magnitude level.  This error does not propagate into the computer codes used to calculate results in \cite{gaillard22}.},
\begin{equation}
\chi_{\rm H_2} = \frac{A_{\rm H_2} f_{\rm H_2}}{P_0}.
\end{equation}
The hydrogen fugacity is $f_{\rm H_2} = \phi_{\rm H_2} P_{\rm H_2}$, where $\phi_{\rm H_2}$ and $P_{\rm H_2}$ are the fugacity coefficient and partial pressure of molecular hydrogen, respectively.  The coefficient relating the mass fraction of hydrogen dissolved in the melt ($\chi_{\rm H_2}$) and the hydrogen fugacity is
\begin{equation}
A_{\rm H_2} = 10^{-2} \exp{\left[-9.43 - \frac{0.181 \mbox{ K}}{T} \left( \frac{P}{P_0} \right) \right]},
\end{equation}
where $P$ is the total atmospheric surface pressure and $T$ is the melt temperature.  The preceding solubility law is an improvement over that used by \cite{kite19}, because it is calibrated on both low and high pressure data.  It is valid for $T \le1400 \mbox{ K}$ and $P \le 3 \mbox{ GPa} = 30 \mbox{ kbar}$.

Figure \ref{fig:solubility} illustrates the relevance of including non-unity fugacity coefficients ($\phi_i \ne 1$) in the solubility laws of both water and hydrogen.  At 1600 K, the discrepancy for the mass fraction of hydrogen dissolved in the melt may be an order of magnitude if one sets $\phi_i=1$.  Generally, the mass fraction of hydrogen dissolved in the melt is small: $\sim 1\%$ at $\sim 10$ kbar.  The mass fraction of water dissolved in melt depends on its mixing ratio, but is generally well below $1\%$ unless the atmosphere becomes water-dominated.  As expected, the solubility depends predominantly on pressure and has a weak dependence on temperature.

\subsubsection{Net chemical reactions and equilibrium constants}

There is some mathematical freedom in how to express a set of chemical reactions needed for calculation.  \cite{french66} wrote down a set of 4 chemical reactions that explicitly involve graphite in a CHO system.  It is possible to rewrite the same set of equations into a pair of equations involving the inter-conversion of carbon dioxide (CO$_2$), carbon monoxide (CO), water (H$_2$O), methane (CH$_4$) and hydrogen (H$_2$) \citep{ht16}.  We also include sulfur dioxide (SO$_2$) and hydrogen sulfide (H$_2$S).  We follow the approach of \cite{gs14}, which explicitly includes oxygen (O$_2$) but excludes graphite:
\begin{equation}
\begin{split}
\mbox{CO} + \frac{1}{2}\mbox{O}_2 &\Longleftrightarrow \mbox{CO}_2, \\
\mbox{H}_2 + \frac{1}{2}\mbox{O}_2 &\Longleftrightarrow \mbox{H}_2\mbox{O}, \\
\mbox{CH}_4 + 2\mbox{O}_2 &\Longleftrightarrow \mbox{CO}_2 + 2\mbox{H}_2\mbox{O}, \\
\frac{1}{2}\mbox{S}_2 + \mbox{O}_2 &\Longleftrightarrow \mbox{SO}_2, \\
\mbox{H}_2\mbox{S} + \frac{1}{2}\mbox{O}_2 &\Longleftrightarrow \frac{1}{2}\mbox{S}_2 + \mbox{H}_2\mbox{O}. \\
\end{split}
\label{eq:CHOS}
\end{equation}
This formulation allows the oxygen fugacity to be directly involved in the equilibrium constants of each of the preceding reactions.  By adding or subtracting various combinations of the equations in (\ref{eq:CHOS}), they may be re-expressed as (e.g., \citealt{ht16})
\begin{equation}
\begin{split}
\mbox{CH}_4 + 2\mbox{H}_2\mbox{O} &\Longleftrightarrow \mbox{CO}_2 + 4\mbox{H}_2, \\
\mbox{CO} + \mbox{H}_2\mbox{O} &\Longleftrightarrow \mbox{H}_2 + \mbox{CO}_2, \\
\mbox{CH}_4 + \mbox{H}_2\mbox{O} &\Longleftrightarrow \mbox{CO} + 3\mbox{H}_2, \\
\end{split}
\end{equation}
which serves as a consistency check.

It has been previously noted that the atmosphere-core interface of sub-Neptunes may reach temperatures exceeding 5000 K, implying that silicate vapour becomes relevant \citep{ms22,charnoz23,ito25}.  Following \cite{misener23}, we include the chemical reaction that converts silicon monoxide (SiO) into silane (SiH$_4$),
\begin{equation}
\mbox{SiO} + 3\mbox{H}_2 \Longleftrightarrow \mbox{SiH}_4 + \mbox{H}_2\mbox{O}. \\
\label{eq:silicon}
\end{equation}

\begin{figure}[!ht]
\begin{center}
\vspace{-0.2in} 
\includegraphics[width=\columnwidth]{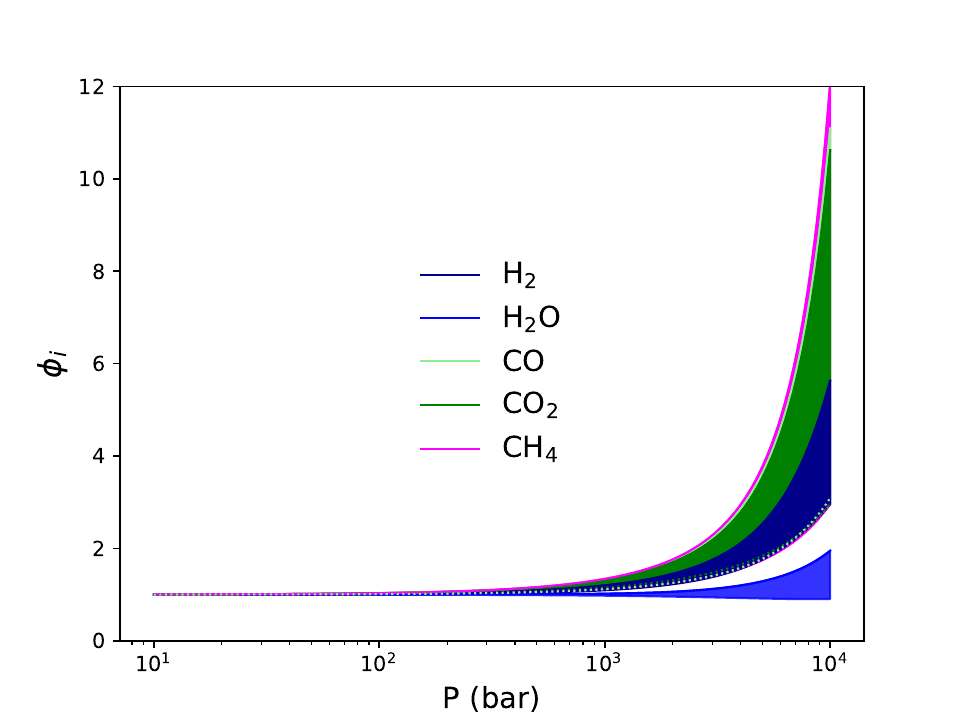}
\end{center}
\caption{Fugacity coefficients for H$_2$, H$_2$O, CO, CO$_2$ and CH$_4$ as functions of pressure.  The shaded region is bounded by calculations for 1600 K (upper bound) and 5000 K (lower bound), corresponding to the temperatures of Earth-like melt and sub-Neptune core surfaces, respectively.  The dotted curves are the lower bounds for CO and CO$_2$, which are otherwise obscured by the shaded regions for H$_2$ and CH$_4$.}
\label{fig:fugacity}
\end{figure}

Using equation (10) of \cite{th24}, the (dimensionless) equilibrium constants of the reactions in equations (\ref{eq:CHOS}) and (\ref{eq:silicon}) may be calculated:
\begin{equation}
\begin{split}
K_1 &= \frac{\alpha_{\rm CO_2} P_{\rm CO_2} P_0^{1/2}}{\alpha_{\rm CO} P_{\rm CO} \left( \gamma_{\rm O_2} f_{\rm O_2} \right)^{1/2} }, \\
K_2 &= \frac{\alpha_{\rm H_2O} P_{\rm H_2O} P_0^{1/2}}{\alpha_{\rm CO} P_{\rm CO} \left( \gamma_{\rm O_2} f_{\rm O_2} \right)^{1/2} }, \\
K_3 &= \frac{\left( \alpha_{\rm H_2O} P_{\rm H_2O} \right)^2 \alpha_{\rm CO_2} P_{\rm CO_2}}{\left( \gamma_{\rm O_2} f_{\rm O_2} \right)^2 \alpha_{\rm CH_4} P_{\rm CH_4}}, \\
K_4 &= \frac{\alpha_{\rm SO_2} P_{\rm SO_2} P_0^{1/2}}{\gamma_{\rm O_2} f_{\rm O_2} \left( \gamma_{\rm S_2} f_{\rm S_2} \right)^{1/2} }, \\
K_5 &= \frac{\alpha_{\rm H_2O} P_{\rm H_2O} \left( \gamma_{\rm S_2} f_{\rm S_2} \right)^{1/2}}{\alpha_{\rm H_2S} P_{\rm H_2S} \left( \gamma_{\rm O_2} f_{\rm O_2} \right)^{1/2} }, \\
K_6 &= \frac{\alpha_{\rm H_2O} \alpha_{\rm SiH_4} P_{\rm H_2O} P_{\rm SiH_4} P^2_0}{\alpha_{\rm SiO} \alpha^3_{\rm H_2} P_{\rm SiO} P^3_{\rm H_2}},
\end{split}
\label{eq:equilibrium_constants}
\end{equation}
where $P_i$ is the partial pressure, $\gamma_i$ is the activity coefficient and $\phi_i$ is the fugacity coefficient of species $i$, respectively.  For convenience, we define $\alpha_i \equiv \gamma_i \phi_i$ following \cite{th24}.  The reference pressure is again $P_0 = 1$ bar.  The equilibrium constants are calculated from Gibbs free energies and are generally functions of temperature and pressure.  In the current study, we set all $\gamma_i=1$ but state them in the derivation for completeness and as a reference for future work.

\subsubsection{Thermodynamic quantities}

The equilibrium constants in equation (\ref{eq:equilibrium_constants}) are related to the difference in Gibbs free energies ($\Delta G$) between the products and reactants via (e.g., \citealt{atkins,devoe}),
\begin{equation}
K_j = \exp{\left( -\frac{\Delta G_j}{{\cal R} T} \right)}
\label{eq:keq}
\end{equation}
where the index $j$ runs from 1 to 6 and ${\cal R}$ is the universal gas constant.  In Appendix \ref{append:gibbs}, we describe how the Gibbs free energies are obtained and provide fitting functions for $\Delta G_j$ corresponding to the 6 net chemical reactions in equations (\ref{eq:CHOS}) and (\ref{eq:silicon}).  Details for how to construct $\Delta G$ for net reactions have previously been provided (e.g., \citealt{hl16}).

Figure \ref{fig:fugacity} shows the fugacity coefficients $\phi_i$ for H$_2$, H$_2$O, CO, CO$_2$ and CH$_4$.  Details for how to calculate the fugacity coefficients are given in Appendix \ref{append:fugacity}.  The fugacity coefficients for the other species are unavailable\footnote{For the sulfur species, the CORK equations of state are unavailable (Appendix \ref{append:fugacity}).  For the silicon species, the equations of state are generally unavailable.}.  For pressures below 1000 bar, $\phi_i \approx 1$ for all of these species.  However, $\phi_i$ begins to significantly depart from unity for pressures of 1000 bar and above, implying that the ideal gas law cannot be assumed for the atmospheres of sub-Neptunes near their cores.

\subsubsection{Total pressure}

The total pressure, which is interpreted as the atmospheric surface pressure \citep{gs14}, is given by
\begin{equation}
\begin{split}
P = &P_{\rm CO} + P_{\rm CO_2} + P_{\rm CH_4} + P_{\rm H_2O} + P_{\rm H_2} + P_{\rm O_2} \\
&+ P_{\rm S_2} + P_{\rm SO_2} + P_{\rm H_2S} + P_{\rm SiO} + P_{\rm SiH_4}.
\end{split}
\label{eq:total_pressure}
\end{equation}
The oxygen ($f_{\rm O_2}$) and sulfur ($f_{\rm S_2}$) fugacities are generalizations of their partial pressures ($P_{\rm O_2}$ and $P_{\rm S_2}$) to allow for departures from an ideal-gas equation of state.  These quantities are related by the fugacity coefficient ($\phi_{\rm O_2}$ and $\phi_{\rm S_2}$) via $f_{\rm O_2} = \phi_{\rm O_2} P_{\rm O_2}$ and $f_{\rm S_2} = \phi_{\rm S_2} P_{\rm S_2}$.  Typical values of $f_{\rm O_2}$ and $f_{\rm S_2}$ for Earth have been reviewed in Section 2.4 of \cite{th24}.  Since we are exploring a wide range of conditions, we do not assume that $P_{\rm O_2}$ and $P_{\rm S_2}$ are negligible in equation (\ref{eq:total_pressure}).  In the absence of data, we assume $\phi_{\rm O_2} = 1$ and $\phi_{\rm S_2} = 1$.

\subsubsection{Reconciling Dalton's law with Newton's second law}

Dalton's law is the statement that the sum of the partial pressures of gases of a mixture is equal to the total pressure of the system (e.g., \citealt{devoe}),
\begin{equation}
P = \sum_i P_i,
\end{equation}
which holds regardless of whether an ideal gas is assumed.

Pressure is the force per unit area.  Written in terms of force per unit area, Newton's second law is
\begin{equation}
P = \frac{M_{\rm atm} g}{4 \pi R_{\rm core}^2},
\end{equation}
where $M_{\rm atm}$ is the mass of the atmosphere, $g$ is the surface gravity of the exoplanet and $R_{\rm core}$ is its radius of the core\footnote{Note that this refers to the entire rocky core of the exoplanet and not just the metallic core as is the convention of the Earth sciences.}.  Let $M_i$ be the mass of species $i$ present in the atmosphere.  By analogy, it is tempting to write a similar expression for the partial pressure, but this comes with the implication that
\begin{equation}
P_i = \frac{M_i g}{4 \pi R^2_{\rm core}} \implies \frac{P_i}{P} = \frac{M_i}{M_{\rm atm}},
\label{eq:partial_pressure_gravity_wrong}
\end{equation}
which is the \textit{mass} mixing ratio rather than the volume mixing ratio.  This is obviously a contradiction to Dalton's law, which implies that $X_i = P_i/P$.

To proceed, we need to clarify a potential source of confusion concerning terminology across different scientific disciplines.  The molecular mass, defined in this study as $m_i$, is simply the mass of a molecule of species $i$.  It has physical units of mass.  If we write the atomic mass unit (amu) as $m_{\rm u}$, then the molecular \textit{weight} is given by $\mu_i  = m_i/m_{\rm u}$.  It is a dimensionless quantity.  A third quantity, the so-called ``molar mass" (used in, e.g., \citealt{bower19}), is the \textit{average} mass of an ensemble of particles of species $i$ and has physical units of mass per mole (g mol$^{-1}$).  All three quantities have their counterparts when averaged over the entire chemical system (containing an arbitrary number of species): mean molecular mass ($\bar{m}$), mean molecular weight ($\mu$) and mean molar mass.  In the current study, we utilize only $m_i$, $\bar{m}$, $\mu_i$ and $\mu$ to minimize confusion.

The solution to this conundrum is to modify equation (\ref{eq:partial_pressure_gravity_wrong}) such that Dalton's law and Newton's second law are in agreement \citep{bower19},
\begin{equation}
P_i = \frac{M_i g}{4 \pi R_{\rm core}^2} \frac{\mu}{\mu_i} = \frac{M_i g}{4 \pi R_{\rm core}^2} \frac{\bar{m}}{m_i}.
\label{eq:partial_pressure_gravity}
\end{equation}
Let $N_i$ be the number of particles of species $i$ and $N$ the total number of particles in the system. Generally, the following relationship holds regardless of whether the system follows an ideal gas law,
\begin{equation}
M_i = m_i N_i = \mu_i m_{\rm u} N_i.
\end{equation}
By applying a summation of equation (\ref{eq:partial_pressure_gravity}),
\begin{equation}
\begin{split}
P &= \sum_i P_i = \sum_i \frac{M_i g}{4 \pi R_{\rm core}^2} \frac{\bar{m}}{m_i} \\
&= \frac{\bar{m} g}{4 \pi R_{\rm core}^2} \sum_i N_i = \frac{M_{\rm atm} g}{4 \pi R_{\rm core}^2},
\end{split}
\end{equation}
since $N = \sum_i N_i$ and $M_{\rm atm} = \bar{m} N$.  Thus, one recovers Newton's second law.  Furthermore, one naturally obtains
\begin{equation}
\frac{P_i}{P} = \frac{M_i}{M_{\rm atm}} \frac{\mu}{\mu_i} = \frac{N_i}{N} = X_i.
\end{equation}

\subsubsection{Hydrogen budget}

To calculate the hydrogen budget of the system, let $N_i^\prime$ and $M_i^\prime$ be the number and mass of species $i$ dissolved in the melt, respectively.  Let $N_{\rm H}$ and $M_{\rm H}$ be the total number and mass of hydrogen atoms of the system, respectively.  By accounting for all of the hydrogen in the system, we obtain
\begin{equation}
\begin{split}
&2 N_{\rm H_2} + 2 N_{\rm H_2}^\prime + 2 N_{\rm H_2O} + 2 N_{\rm H_2O}^\prime \\
&+ 4 N_{\rm CH_4} + 2 N_{\rm H_2S} + 4 N_{\rm SiH_4} = N_{\rm H},
\end{split}
\end{equation}
where we have approximated $N_{\rm CH_4}^\prime \approx 0$, $N_{\rm H_2S}^\prime \approx 0$ and $N_{\rm SiH_4}^\prime \approx 0$.  Multiplying the preceding equation by $m_{\rm H}$ (the mass of the hydrogen atom), we obtain
\begin{equation}
\begin{split}
&2 m_{\rm H} \left( N_{\rm H_2} + N_{\rm H_2}^\prime \right) + \frac{2}{\mu_{\rm H_2O}} m_{\rm H_2O} \left( N_{\rm H_2O} + N_{\rm H_2O}^\prime \right) \\
&+ \frac{4}{\mu_{\rm CH_4}} m_{\rm CH_4} N_{\rm CH_4} + \frac{2}{\mu_{\rm H_2S}} m_{\rm H_2S} N_{\rm H_2S} \\
&+ \frac{4}{\mu_{\rm SiH_4}} m_{\rm SiH_4} N_{\rm SiH_4} = m_{\rm H} N_{\rm H},
\end{split}
\end{equation}
where $m_i$ is the mass of species $i$ and we have approximated $m_{\rm H} \approx m_{\rm u}$.  Since $m_i N_i = M_i$ and $m_i N_i^\prime = M_i^\prime$, it follows that 
\begin{equation}
\begin{split}
&M_{\rm H_2} + M_{\rm H_2}^\prime + \frac{2}{\mu_{\rm H_2O}} \left( M_{\rm H_2O} + M_{\rm H_2O}^\prime \right) + \frac{4}{\mu_{\rm CH_4}} M_{\rm CH_4} \\
&+ \frac{2}{\mu_{\rm H_2S}} M_{\rm H_2S} + \frac{4}{\mu_{\rm SiH_4}} M_{\rm SiH_4} = M_{\rm H}.
\end{split}
\end{equation}

Let the total mass of melt participating in the outgassing be $M_{\rm melt}$.  It follows that 
\begin{equation}
M_i^\prime = \chi_i M_{\rm melt},
\end{equation}
where the mass fraction of species $i$ dissolved in the melt is given by the solubility laws of Section \ref{subsect:solubility}.  For algebraic convenience, we rewrite these solubility laws as
\begin{equation}
\begin{split}
\chi_{\rm H_2O} &= A^\prime_{\rm H_2O} P_{\rm H_2O}^{1/2}, \\
\chi_{\rm H_2} &= A^\prime_{\rm H_2} P_{\rm H_2}, \\
\end{split}
\end{equation}
where the coefficients have been rewritten as
\begin{equation}
\begin{split}
A^\prime_{\rm H_2O} &\equiv A_{\rm H_2O} \left(\frac{\phi_{\rm H_2O}}{P_0} \right)^{1/2}, \\
A^\prime_{\rm H_2} &\equiv \frac{A_{\rm H_2} \phi_{\rm H_2}}{P_0}. \\
\end{split}
\end{equation}

The hydrogen budget may be expressed in terms of gaseous partial pressures ($P_i$) and the total pressure
\begin{equation}
\begin{split}
&P_{\rm H_2} \left( 2 + \mu P \chi_{\rm melt} A^\prime_{\rm H_2} \right) + \frac{1}{9} \mu P \chi_{\rm melt} A^\prime_{\rm H_2O} P_{\rm H_2O}^{1/2} \\
&+ 2 P_{\rm H_2O} + 4 P_{\rm CH_4} + 2 P_{\rm H_2S} + 4 P_{\rm SiH_4}= P_{\rm H},
\end{split}
\label{eq:mass_budget}
\end{equation}
where $\chi_{\rm melt} \equiv M_{\rm melt}/M_{\rm atm}$ is the ratio of the total mass of the melt to the total atmospheric mass.

\subsubsection{Carbon, oxygen and silicon budgets}

Since the solubility of carbon species in melt is not considered in the current study, the carbon budget becomes trivial to write down:
\begin{equation}
N_{\rm CO} + N_{\rm CO_2} + N_{\rm CH_4} = N_{\rm C},
\end{equation}
where $N_{\rm C}$ is the total number of carbon atoms in the system.  By multiplying the preceding equation by $m_{\rm C} g \mu / 4 \pi R^2_{\rm core}$ and using equation (\ref{eq:partial_pressure_gravity}), we obtain
\begin{equation}
\frac{\mu_{\rm C} \mu g}{4 \pi R^2_{\rm core}} \left( \frac{M_{\rm CO}}{\mu_{\rm CO}} + \frac{M_{\rm CO_2}}{\mu_{\rm CO_2}} + \frac{M_{\rm CH_4}}{\mu_{\rm CH_4}} \right) = \frac{M_{\rm C} g \mu}{4 \pi R^2_{\rm core}},
\end{equation}
where $M_{\rm C} = m_{\rm C} N_{\rm C}$ is the total mass of carbon atoms in the system.  The carbon budget is thus expressed in terms of partial pressures,
\begin{equation}
P_{\rm CO} + P_{\rm CO_2} + P_{\rm CH_4} = \frac{M_{\rm C} g}{4 \pi R^2_{\rm core}} \frac{\mu}{\mu_{\rm C}} = P_{\rm C}.
\end{equation}
The preceding result implies that, when solubility is ignored, conserving the total number of carbon atoms in a system may be expressed in terms of the individual partial pressures of molecules.

Similarly, the oxygen and silicon budgets may be straightforwardly written down:
\begin{equation}
\begin{split}
&P_{\rm CO} + 2 P_{\rm CO_2} + P_{\rm H_2O} + 2 P_{\rm SO_2} + P_{\rm SiO} + 2 P_{\rm O_2} = P_{\rm O}, \\
&P_{\rm SiO} + P_{\rm SiH_4} = P_{\rm Si}.
\end{split}
\end{equation}
It is worth noting that $P_{\rm O}$ and $P_{\rm Si}$ include only silicon and oxygen in the atmosphere (gas phase), an assumption that will be justified in Section \ref{subsect:model_design}.

We note that $P_{\rm C}/P_{\rm H} = N_{\rm C}/N_{\rm H}$ and $P_{\rm Si}/P_{\rm O} = N_{\rm Si}/N_{\rm O}$.

\subsection{Solution method}

The equilibrium constants in equation (\ref{eq:equilibrium_constants}) allow us to write down the following relationships between the partial pressures of molecules:
\begin{equation}
\begin{split}
P_{\rm CO_2} &= F_1 P_{\rm CO}, \\
P_{\rm H_2O} &= F_2 P_{\rm H_2}, \\
P_{\rm CH_4} &= F_3 P_{\rm CO_2} P_{\rm H_2O}^2 = F_4 P_{\rm CO} P_{\rm H_2}^2, \\
P_{\rm H_2S} &= F_7 P_{\rm H_2O}, \\
P_{\rm SiH_4} &= \frac{F_8 P_{\rm SiO} P_{\rm H_2}^3}{P_{\rm H_2O}}.
\end{split}
\label{eq:partial_F}
\end{equation}
The carbon, hydrogen, silicon and oxygen budgets may be combined to form a pair of equations for the carbon-to-hydrogen and silicon-to-oxygen ratios,
\begin{equation}
\begin{split}
&P_{\rm CO} + P_{\rm CO_2} + P_{\rm CH_4} = \frac{N_{\rm C}}{N_{\rm H}} \left[ P_{\rm H_2} \left( 2 + \mu F_5 P \right) + 2P_{\rm H_2O} \right.\\
&\left.+ \frac{1}{9} \mu F_6 P P_{\rm H_2}^{1/2} + 4 P_{\rm CH_4} + 2 P_{\rm H_2S} + 4 P_{\rm SiH_4} \right] , \\
&P_{\rm SiO} + P_{\rm SiH_4} = \frac{N_{\rm Si}}{N_{\rm O}} \left( P_{\rm CO} + 2 P_{\rm CO_2} + P_{\rm H_2O} + 2 P_{\rm SO_2} \right.\\
&\left.+ P_{\rm SiO} + 2 P_{\rm O_2} \right),
\end{split}
\label{eq:elemental_budgets}
\end{equation}
where $\mbox{C/H} \equiv N_{\rm C}/N_{\rm H}$ is the elemental abundance of carbon (relative to hydrogen).  Its solar value is $\mbox{C/H} = 2.5 \times 10^{-4}$.  The ratio of silicon to oxygen (by number) is given by $\mbox{Si/O} \equiv N_{\rm Si}/N_{\rm O}$.  For example, \cite{misener23} considered the conversion of gaseous silica (liquid SiO$_2$) to silicon monoxide and assumed $\mbox{Si/O}=0.5$ for their chemical system.

The various coefficients in the preceding equations are
\begin{equation}
\begin{split}
F_1 &\equiv \frac{K_1 \alpha_{\rm CO}}{\alpha_{\rm CO_2}} \left( \frac{\gamma_{\rm O_2} f_{\rm O_2}}{P_0} \right)^{1/2}, \\
F_2 &\equiv \frac{K_2 \alpha_{\rm H_2}}{\alpha_{\rm H_2O}} \left( \frac{\gamma_{\rm O_2} f_{\rm O_2}}{P_0} \right)^{1/2}, \\
F_3 &\equiv \frac{\alpha_{\rm CO_2}}{K_3 \alpha_{\rm CH_4}} \left( \frac{\alpha_{\rm H_2O}}{\gamma_{\rm O_2} f_{\rm O_2}} \right)^2, \\
F_4 &\equiv F_1 F_2^2 F_3, \\
F_5 &\equiv \chi_{\rm melt} A^\prime_{\rm H_2}, \\
F_6 &\equiv \chi_{\rm melt} A^\prime_{\rm H_2O} F^{1/2}_2,\\
F_7 &\equiv \frac{\alpha_{\rm H_2O}}{K_5 \alpha_{\rm H_2S}} \left( \frac{\gamma_{\rm S_2} f_{\rm S_2}}{\gamma_{\rm O_2} f_{\rm O_2}} \right)^{1/2}, \\
F_8 &\equiv \frac{K_6 \alpha_{\rm SiO} \alpha_{\rm H_2}^3}{\alpha_{\rm H_2O} \alpha_{\rm SiH_4} P_0^2}.
\end{split}
\end{equation}
The partial pressure of sulfur dioxide may be straightforwardly calculated once $f_{\rm O_2}$ and $f_{\rm S_2}$ are specified,
\begin{equation}
P_{\rm SO_2} = \frac{K_4 \gamma_{\rm O_2} f_{\rm O_2}}{\alpha_{\rm SO_2}} \left( \frac{\gamma_{\rm S_2} f_{\rm S_2}}{P_0} \right)^{1/2}.
\label{eq:SO2}
\end{equation}

Together with equation (\ref{eq:total_pressure}), which we rewrite as
\begin{equation}
\sum_i X_i = 1,
\end{equation}
where $X_i \equiv P_i/P$, and
\begin{equation}
\begin{split}
\mu =& 2 X_{\rm H_2} + 16X_{\rm CH_4} + 18 X_{\rm H_2O} + 28 X_{\rm CO} + 32 X_{\rm SiH_4} \\
&+ 34 X_{\rm H_2S} + 44 X_{\rm CO_2} + 44 X_{\rm SiO} + 64 X_{\rm SO_2} \\\
&+ 32 X_{\rm O_2} + 64 X_{\rm S_2}, \\
\end{split}
\end{equation}
the equations in (\ref{eq:partial_F}) and (\ref{eq:elemental_budgets}) form a set of 9 coupled algebraic equations that may be solved using the \texttt{fsolve} module\footnote{The default tolerance ($\sim 10^{-8}$) was used, as well as the default option to estimate the Jacobian numerically.} that is part of the \texttt{scipy} package of the \texttt{Python} programming language.  The solutions of these coupled non-linear algebraic equations may easily be trapped in local minima and not converge.  A crucial ingredient are plausible first guesses for the partial pressures, but such guesses are non-trivial to make as $P_i$ generally spans more than 30 orders of magnitude.  We estimate first guesses using a simplified CHOSSi solution described in Appendix \ref{append:CHOSSi}.  The crucial aspect of this solution is that it is not only analytical, but provides an \textit{explicit} expression for $P$ that enables some of the partial pressures and the mean molecular weight to be computed explicitly.  In other words, the simplified CHOSSi solution does not involve a numerical root-finding step (and is thus unconditionally stable), which allows us to explore a large range of oxygen and sulfur fugacities.  Without such a simplified CHOSSi solution to guide the \texttt{fsolve} module, the numerical solutions suffer from instabilities when attempting to compute $X_i$ across temperature and pressure (not shown).

While we have neglected helium, its addition constitutes only a minor correction ($\sim 0.4$) to the mean molecular weight.

\section{Results}
\label{sect:results}

\subsection{Data-driven radius valley constraints}
\label{subsect:radius_valley}

\begin{figure*}[!ht]
\begin{center}
\vspace{-0.2in} 
\includegraphics[width=\columnwidth]{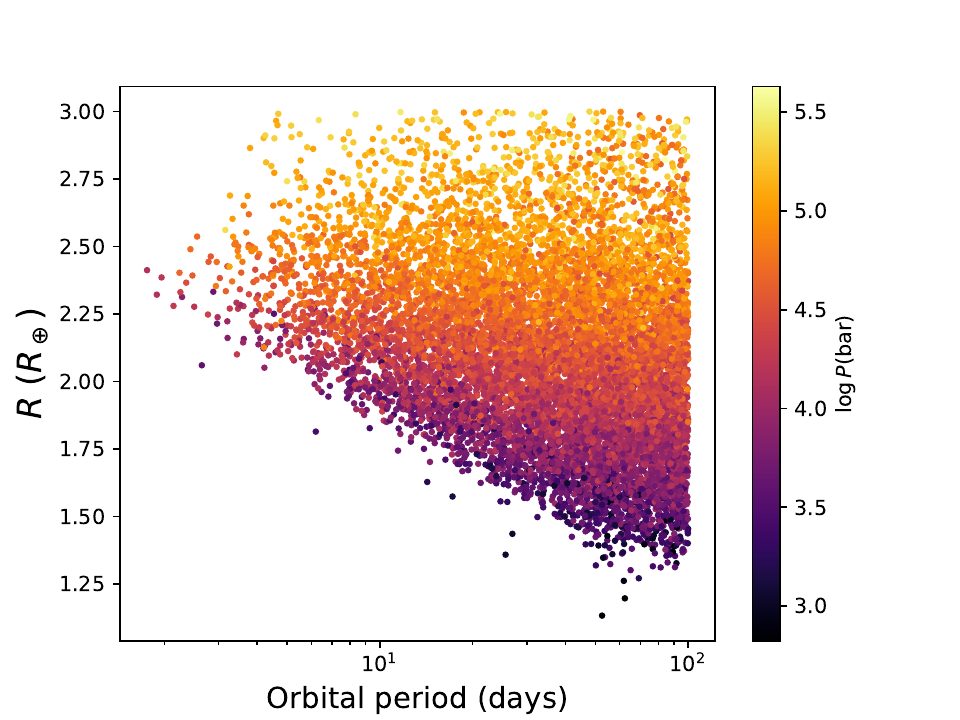}
\includegraphics[width=\columnwidth]{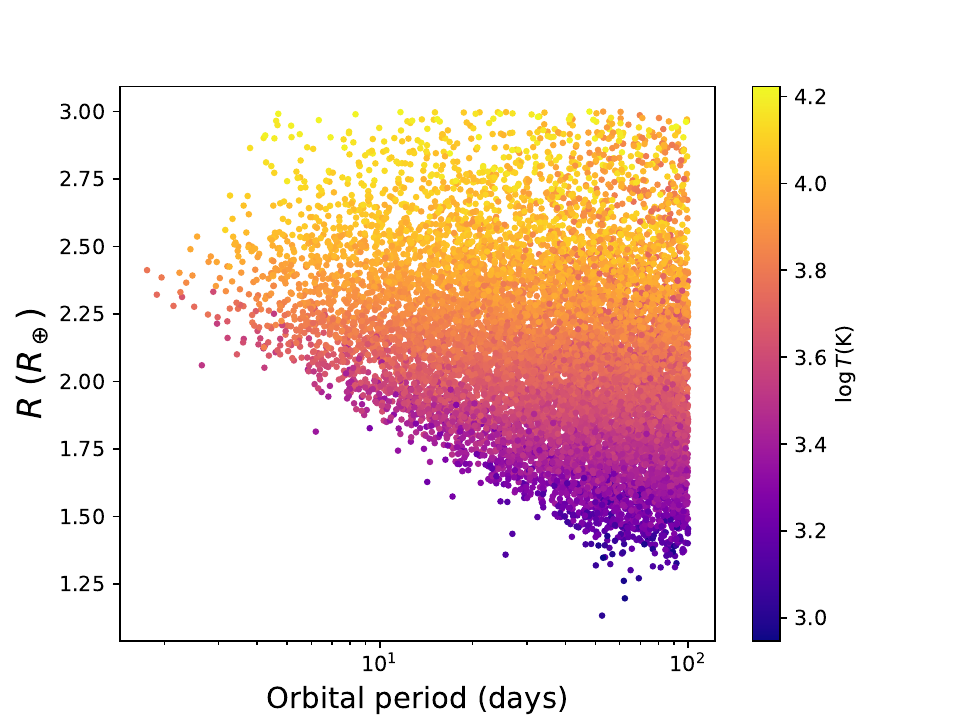}
\includegraphics[width=\columnwidth]{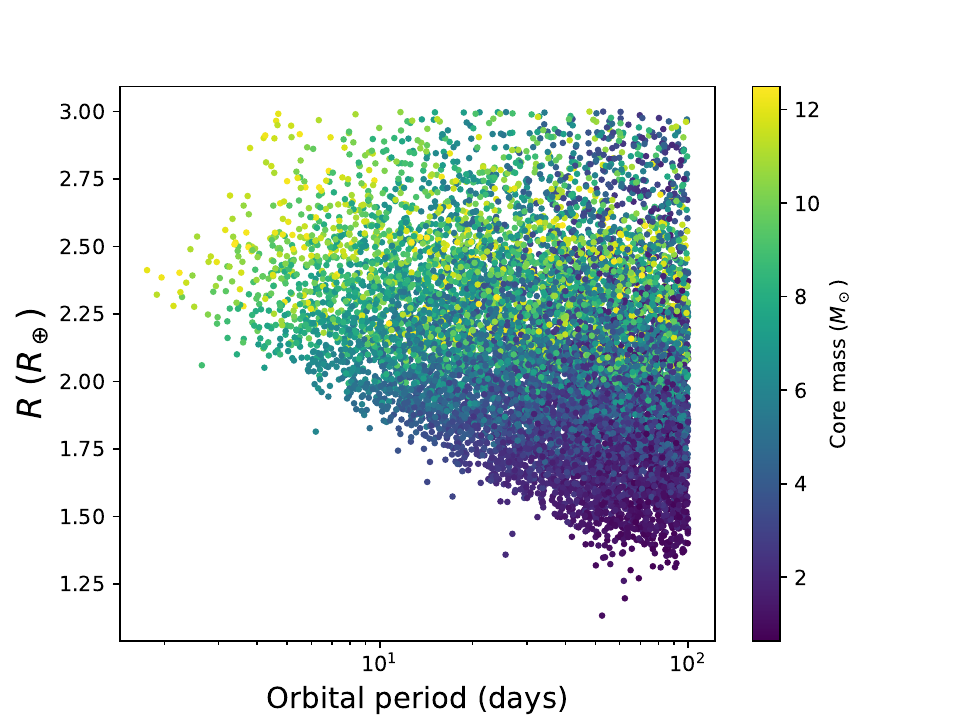}
\includegraphics[width=\columnwidth]{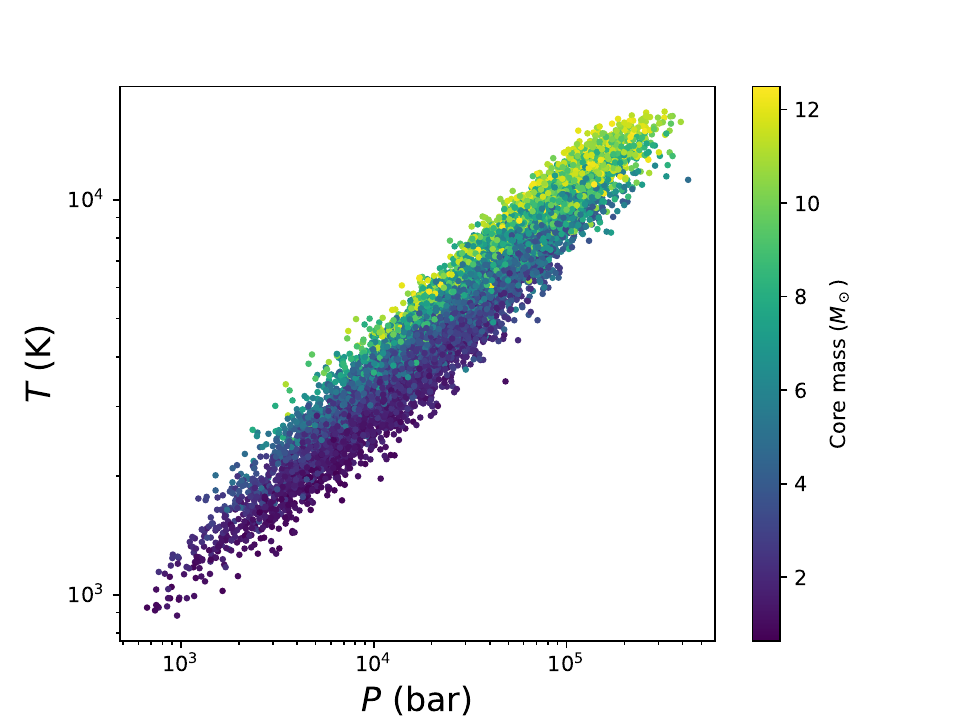}
\end{center}
\caption{Data-driven radius valley constraints from matching the observed distributions of planetary radius and orbital period, based on and extending calculations from \cite{ro21}.  The top-left, top-right and bottom-left panels show the logarithm of the surface pressure (in bar), logarithm of the melt temperature (in K) and core mass (in Earth masses), respectively, as a color scale.  The bottom-right panel shows the temperature-pressure relationship with the core mass as a color scale.}
\label{fig:radius_valley}
\end{figure*}

\cite{ro21} previously modeled the distributions of planetary radius and orbital period across the radius valley and extracted the distributions of atmospheric mass fractions ($M_{\rm atm}/M_{\rm core}$) and core masses ($M_{\rm core}$).  Inherent in these calculations are the temperature-pressure conditions at the atmosphere-core interface, but these calculations were not explicitly reported in that study.

In Figure \ref{fig:radius_valley}, we show these calculations for systems that are 3 Gyr old.  Since photo-evaporation mostly occurs in the first $\sim 100$ Myr of a star's life, the outcomes are insensitive to the age assumption of 3 Gyr and thermal contraction has slowed significantly.  The surface pressure generally decreases with planetary radius, because less massive exoplanets have weaker gravities and thus less massive atmospheres.  They are in the range $\sim 10^3$--$10^5$ bar for $R<3R_\oplus$.  A less massive atmosphere results in more rapid cooling, which produces lower surface temperatures.  The surface temperatures are in the range $\sim 10^3$--$10^4$ K.  When $P \approx 1$ kbar, the surface temperature is about 3000 K.

We will use the surface pressures and temperatures in Figure \ref{fig:radius_valley} to inform $P_{\rm H_2}$ and $T$, respectively, in the outgassing calculations.

\subsection{Model design and choice of parameters}
\label{subsect:model_design}

One of the lessons learned in the current study is the relevance of model design: how the equations, unknowns and parameters are defined.  There are an arbitrary number of ways to formulate this.  Instead of parametrizing the bulk carbon, hydrogen, carbon and silicon content of an exoplanet (which is generally unknown), we describe it by ratios of elemental abundances representing either the atmosphere or the melt (that outgasses the atmosphere). We focus exclusively on hybrid atmospheres and designate H$_2$O, CO, CO$_2$, CH$_4$, H$_2$S, SO$_2$, SiO and SiH$_4$ as the unknown gaseous species whose abundances we wish to solve for.  Since there are 5 net chemical reactions, an equation for C/H, an equation for Si/O, the condition that the partial pressures must sum up to the total pressure and the expression for the mean molecular weight, this formally constitutes 9 equations and 9 unknowns.

Our choice of the following 7 parameters are justified on physical and/or chemical grounds.
\begin{enumerate}

\item \textbf{Hydrogen partial pressure ($P_{\rm H_2}$):} The primordial hydrogen-dominated envelope left over from formation may be quantified by its partial pressure.  In the current study, we are guided by the data-driven inference approach of \cite{ro21}, who derived atmospheric mass fractions that are consistent with the observed radius valley separating super Earths and sub-Neptunes.  These mass fractions may be straightforwardly converted into partial pressures by assuming hydrogen-dominated atmospheres.

\item \textbf{Melt temperature ($T$):} The Gibbs free energies involved in our equilibrium chemistry calculations require a temperature to be specified.  In the current study, this temperature is interpreted as the melt temperature \citep{gs14,gaillard22,th24}.  For super Earths with a solid rocky surface, outgassing occurs from sub-surface melt located in the mantle.  For sub-Neptunes, where the surface of the rocky core may attain temperatures exceeding the melting point (liquidus) of rock \citep{ms22}, the temperature refers to that of the molten magma ocean at the surface of the core.  The core surface temperatures were previously calculated by \cite{ro21} and we use them as input for our outgassing calculations.

\item \textbf{Oxygen fugacity ($f_{\rm O_2}$):} The oxygen fugacity is the effective partial pressure of oxygen gas participating in equilibrium reactions involving multi-valent elements like iron in the melt.  In other words, the oxygen fugacity is \textit{buffered} by the melt.  It is typically specified relative to a chemical buffer, because the amount of gaseous oxygen liberated depends on temperature.  A commonly used buffer is that of iron-w\"{u}stite (IW) \citep{ww05},
\begin{equation}
\mbox{Fe} + \frac{1}{2}\mbox{O}_2 \Longleftrightarrow \mbox{FeO}.
\end{equation}
The upper mantle of the Earth is estimated to have an oxygen fugacity of $\log{f_{\rm O_2}}=\mbox{IW}+3.5$ \citep{fm08}, which corresponds to $f_{\rm O_2} \sim 10^{-7}$ bar for $T=1600$ K and $P=1$ bar \citep{ballhaus91,hirschmann21}.  

Strictly speaking, the approach of specifying the oxygen fugacity relative to the IW buffer cannot be implemented using existing empirical formulae, because they are not calibrated for temperatures above 3000 K \citep{ballhaus91,hirschmann21} (see Appendix \ref{append:IW}).  As a striking example, if IW$+3.5$ is extrapolated to 5000 K using the empirical formula of \cite{hirschmann21}, then it corresponds to $f_{\rm O_2} \approx 28.5$ kbar for a pressure of $10$ kbar, i.e., the amount of gaseous oxygen liberated exceeds the atmospheric surface pressure.  

\begin{figure}[!ht]
\begin{center}
\includegraphics[width=\columnwidth]{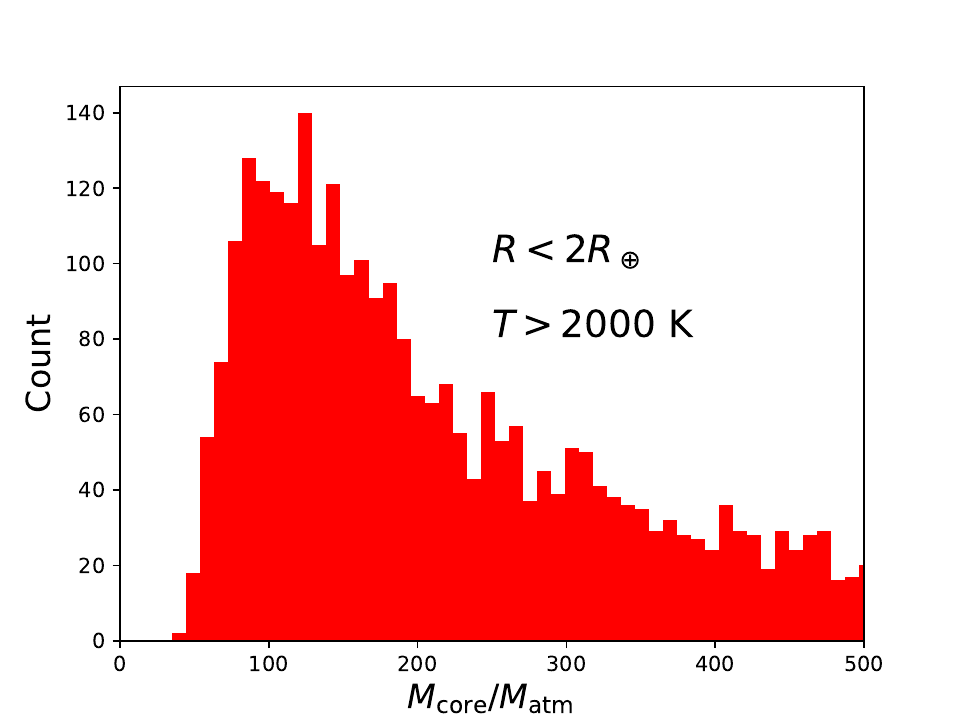}
\end{center}
\vspace{-0.1in}
\caption{Distribution of $M_{\rm core}/M_{\rm atm}$ (the reciprocal of the atmospheric mass fraction), where we have included only models with $R< 2R_\oplus$, a non-zero atmospheric mass and a melt temperature of $T>2000$ K.  This curated sample corresponds to 3758 simulated exoplanets.  The maximum value of $M_{\rm core}/M_{\rm atm}$ in this distribution is about 9462.}
\label{fig:Mratio}
\end{figure}

\begin{figure*}[!ht]
\begin{center}
\vspace{-0.2in} 
\includegraphics[width=\columnwidth]{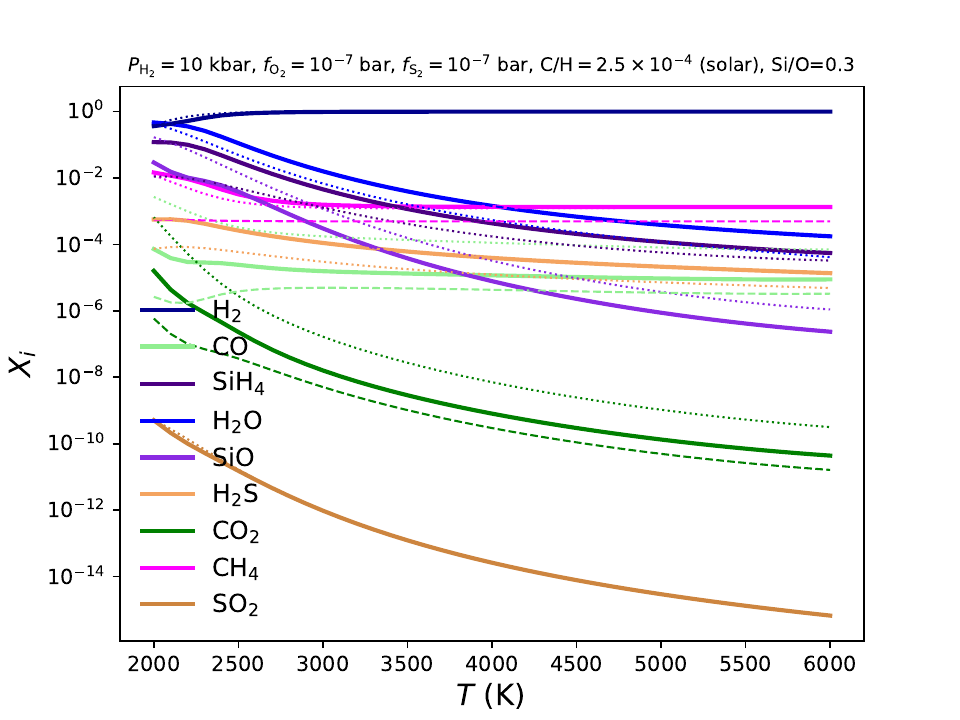}
\includegraphics[width=\columnwidth]{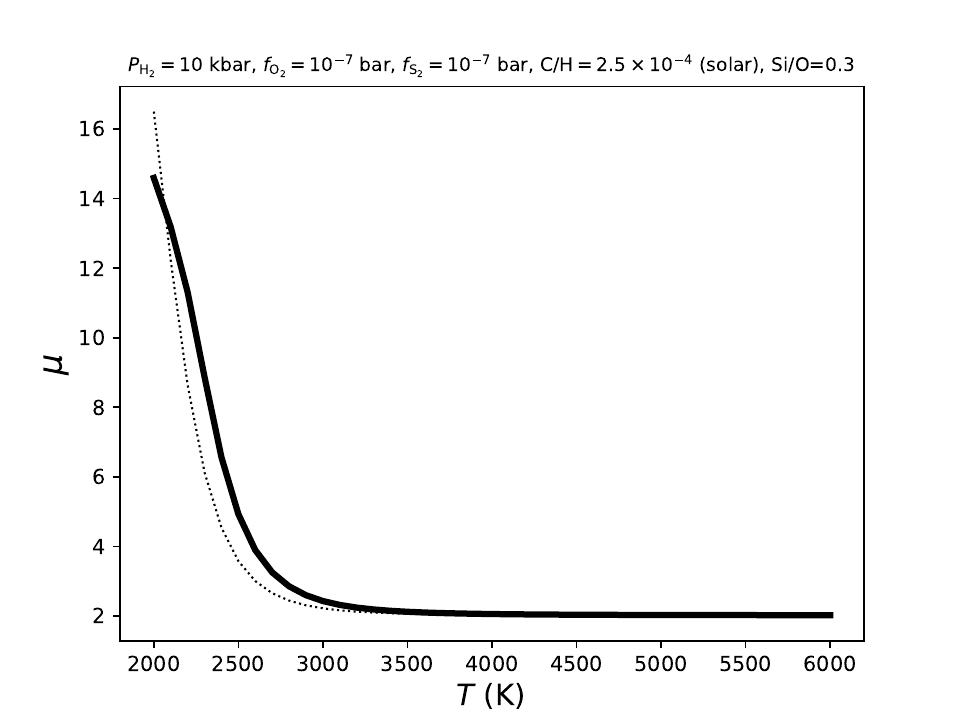}
\includegraphics[width=\columnwidth]{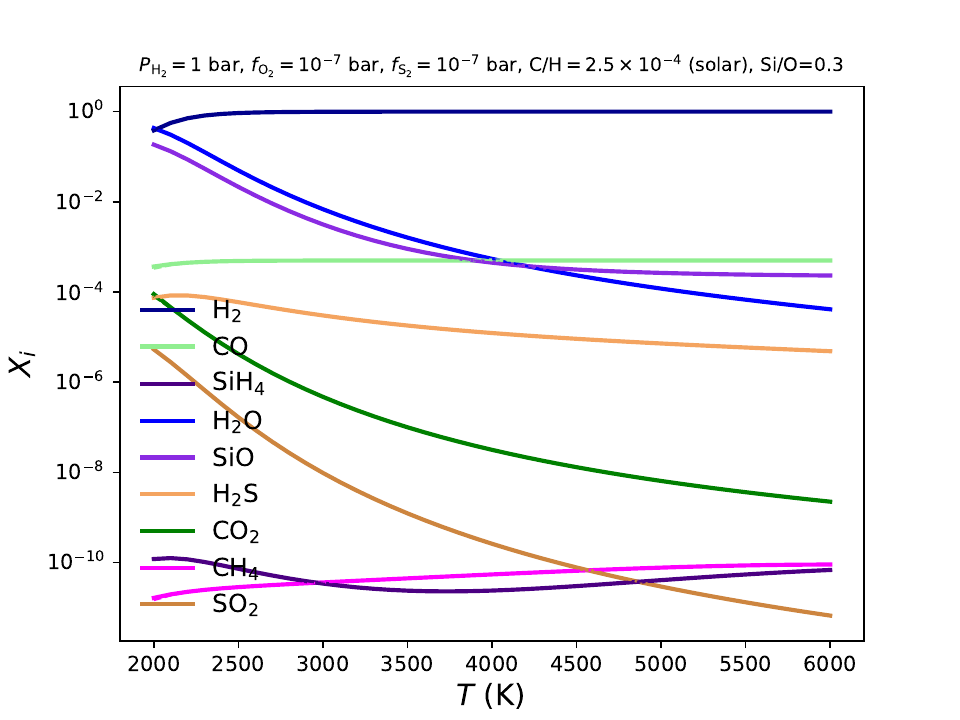}
\includegraphics[width=\columnwidth]{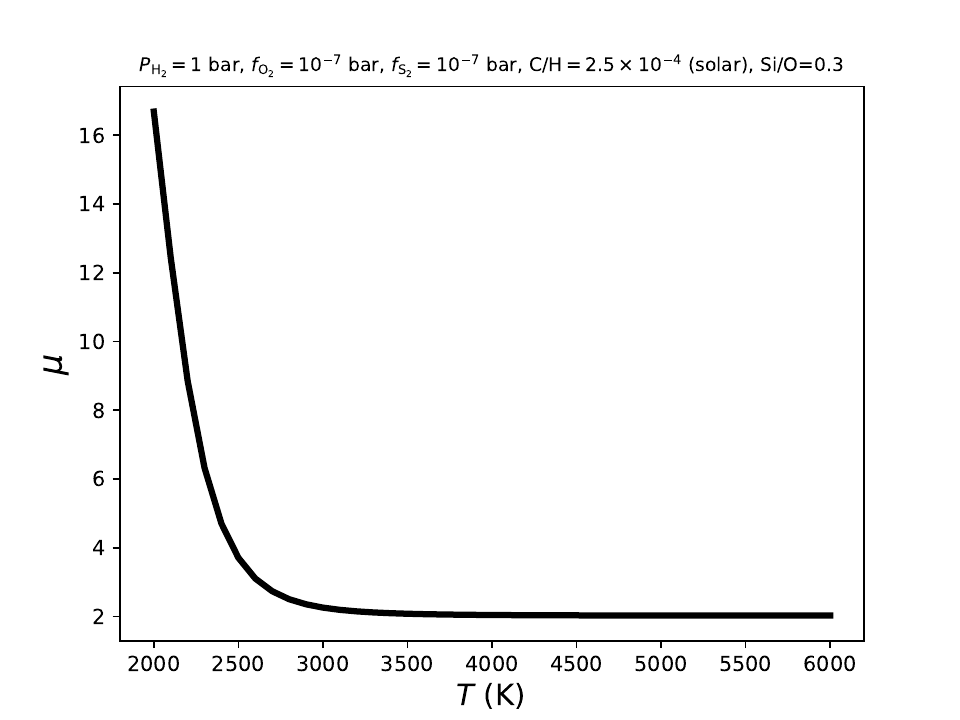}
\end{center}
\vspace{-0.1in}
\caption{Outgassing calculations for sub-Neptune-like ($P_{\rm H_2}=10$ kbar; top row) and Earth-like ($P_{\rm H_2}=1$ bar; bottom row) surface pressures.  For illustration, we have assumed the oxygen and sulfur fugacities to be equal and have the value $f_{\rm O_2} = f_{\rm S_2} = 10^{-1}$ dyne cm$^{-2}$ $=10^{-7}$ bar.  The elemental carbon abundance is set to its solar value ($\mbox{C/H} = 2.5 \times 10^{-4}$), while the silicon-to-oxygen ratio is set to $\mbox{Si/O}=0.3$.  The melt mass fraction has been set to $\chi_{\rm melt}=100$.  The dotted curves are additional calculations that assume ideal gas behavior ($\phi_i=1$) wherever possible (for H$_2$, H$_2$O, CO, CO$_2$ and CH$_4$), while the dashed curves ignore the solubility laws (for H$_2$ and H$_2$O).  Left column: volume mixing ratios of various molecules as functions of melt temperature.  Non-ideal gas behavior is only relevant for the sub-Neptune-like surface pressure.  Right column: mean molecular weight as a function of melt temperature.  For the sub-Neptune-like case study, the $\log{f_{\rm O_2}}-\mbox{IW}$ values range from $-8.7$ to $-0.2$ across the temperature range of $2000$--$6000$ K.  For the Earth-like case study, it ranges from $-8.5$ to $0.1$.}
\vspace{0.1in}
\label{fig:examples}
\end{figure*}

The Solar System gives no guidance on how to predict the oxygen fugacity of a rocky body with a mass exceeding that of Earth.  If we use the estimated values for the upper mantles of Earth (IW$+3.5$) and Mars (IW) \citep{deng20}, then a crude, empirical scaling relationship may be obtained,
\begin{equation}
\log{f_{\rm O_2}} = \mbox{IW} + \frac{3.921 M}{M_\oplus} - 0.421.
\end{equation}
For a rocky core of 4 Earth masses, this scaling relationship produces an implausible value of IW$+15.3$, which corresponds to about $70$ kbar at 1600 K.

A predictive theory for the formation of sub-Neptunes would elucidate how $\log{f_{\rm O_2}}-\mbox{IW}$ behaves as a function of the age, mass, radius, orbital period and the other properties of a sub-Neptune---without hiding our ignorance behind dozens of free parameters.  A survey of the literature reveals that such a predictive theory remains elusive.  For example, \cite{lich21} sketches a qualitative scheme for describing the oxidation state of the mantle, but uses order-of-magnitude scaling relations, which lack predictive power, to describe the unsolved problem of atmospheric mixing (see Section \ref{subsect:mixing}).  As another example, \cite{kite20} considers end members for the core and magma ocean compositions of sub-Neptunes.  In their Section 4.2, \cite{kite20} ruminates: ``Sub-Neptune formation is not well understood.  We do not know how rocky cores form"; ``We do not know where the growth happened"; ``We do not know where the volatiles came from---nebula gas or solid-derived volatiles?"

In the absence of a first-principles theory for the oxygen fugacity of rocky bodies with masses exceeding that of Earth's, we parameterise the oxygen fugacity by its absolute value across a broad range of values: $f_{\rm O_2} = 10^{-11}$--$10^2$ dyne cm$^{-2} = 10^{-17}$--$10^{-4}$ bar.  We will see that interesting transitions in the mean molecular weight occur for $f_{\rm O_2} \gtrsim 10^{-11}$ bar.

\item \textbf{Sulfur fugacity ($f_{\rm S_2}$):} The sulfur fugacity is the amount of gaseous sulfur buffered by the melt.  Its range of values is poorly known even for Earth (see discussion in Section 2.4 of \citealt{th24}).  In the absence of better knowledge or first-principles theory, we parametrize the sulfur fugacity across the same, broad range of values as for the oxygen fugacity: $f_{\rm S_2} = 10^{-11}$--$10^2$ dyne cm$^{-2} = 10^{-17}$--$10^{-4}$ bar.

\item \textbf{Elemental carbon abundance (C/H):} Carbon tends to reside in the gaseous phase rather than be dissolved in the melt.  To lowest order, the carbon-to-hydrogen ratio (by number) describes the enrichment of the primordial hydrogen envelope.  Its solar-abundance value is $\mbox{C/H} = 2.5 \times 10^{-4}$, which we use as a plausible starting point.

\item \textbf{Silicon-to-oxygen ratio of melt (Si/O):} The atmosphere is assumed to have the same Si/O ratio as the melt.  If a pure silica (SiO$_2$) melt is assumed, then one obtains $\mbox{Si/O}=0.5$ \citep{misener23}.  A more realistic basaltic melt contains oxides of silicon, iron, magnesium, etc, as well as dissolved water (cf. Table 1 on pg. 145 of \citealt{winter13}), but a negligible amount of dissolved molecular hydrogen.  We estimate a lower bound of $\mbox{Si/O} \gtrsim 0.3$ based on Mid-Atlantic Ridge basalts \citep{winter13}.  In the current study, we fix $\mbox{Si/O}=0.3$.

\item \textbf{Mass fraction of melt ($\chi_{\rm melt}$):} The melt mass fraction may be rewritten as
\begin{equation}
\chi_{\rm melt} = \frac{M_{\rm melt}}{M_{\rm core}} \left( \frac{M_{\rm atm}}{M_{\rm core}} \right)^{-1}.
\label{eq:chi_melt}
\end{equation}
The atmospheric mass fraction ($M_{\rm atm}/M_{\rm core}$) is provided by radius valley constraints \citep{ro21}.  For hydrogen-dominated sub-Neptunes, we have $M_{\rm atm}/M_{\rm core} \sim 0.01$.  Therefore, a rough estimate for the melt mass fraction is $\chi_{\rm melt} \sim 100$ if $M_{\rm melt}/M_{\rm core} \sim 1$.  In Figure \ref{fig:Mratio}, we show the distribution of $M_{\rm core}/M_{\rm atm}$ derived from the calculations of \cite{ro21}.  The fraction of the sub-Neptune core that is participating in the solubility of gases is difficult to estimate without a more careful treatment of interior geodynamics, which is beyond the scope of the current study.

\end{enumerate}

Effectively, the main parameters are the oxygen and sulfur fugacity, because they may span an enormous range of values and we currently do not have any theory to constrain them.  Furthermore, $f_{\rm O_2}$, $f_{\rm S_2}$ and $T$ may be related by melt chemistry, which we do not treat in the current study.

\begin{figure*}[!ht]
\begin{center}
\vspace{-0.2in} 
\includegraphics[width=\columnwidth]{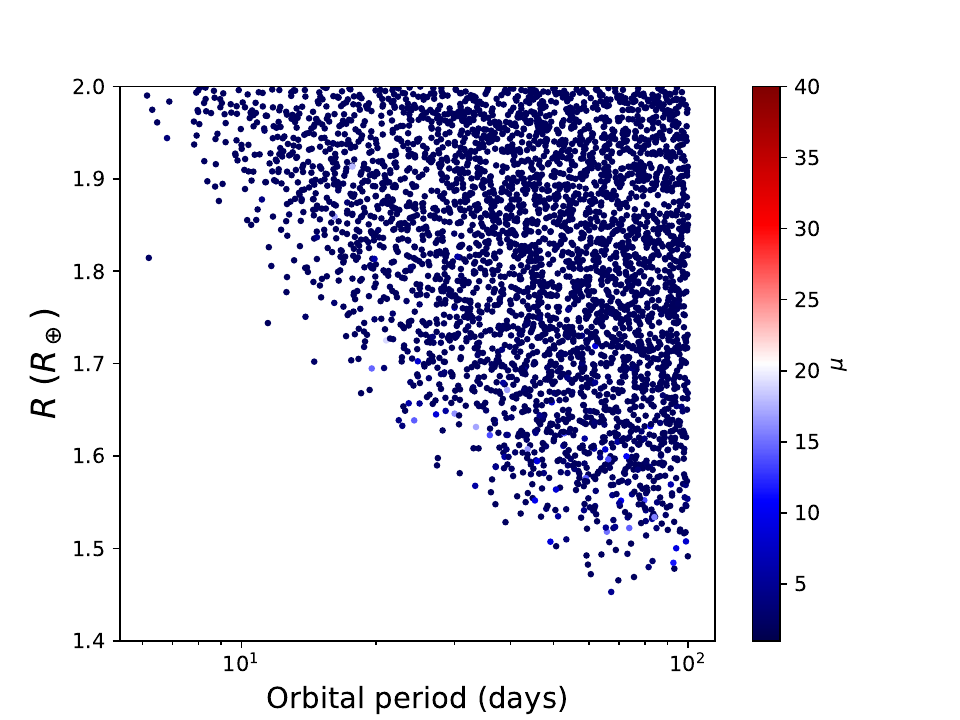}
\includegraphics[width=\columnwidth]{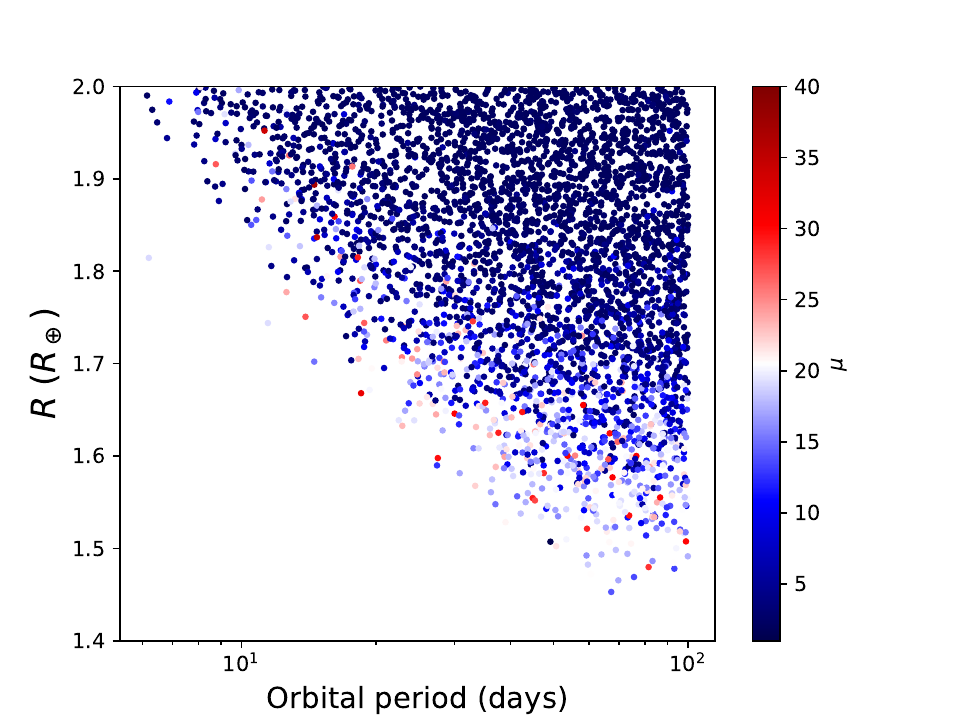}
\includegraphics[width=\columnwidth]{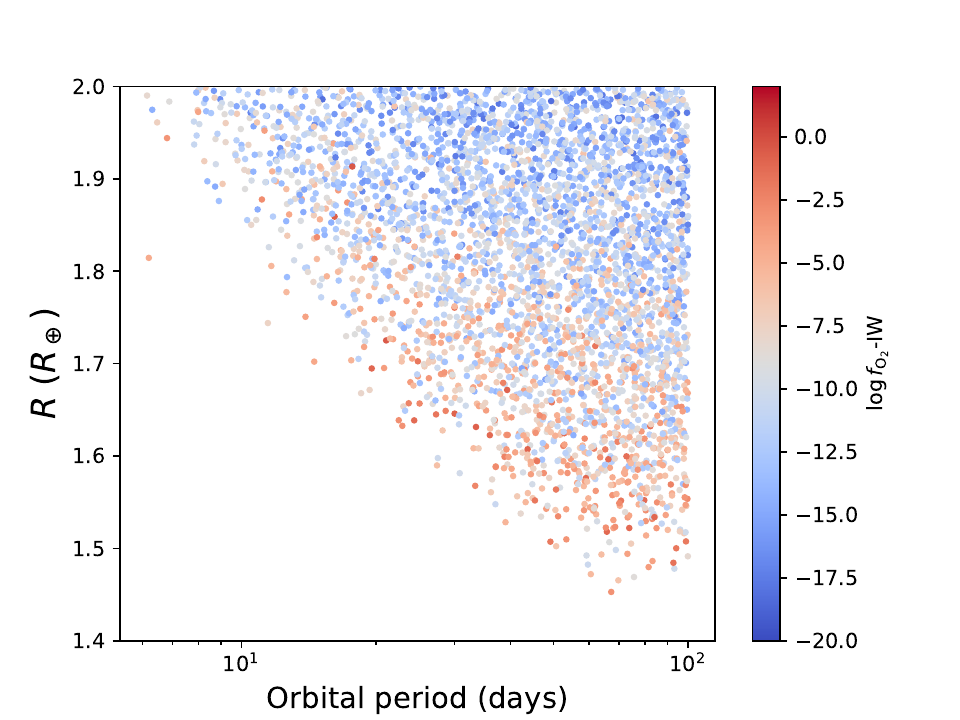}
\includegraphics[width=\columnwidth]{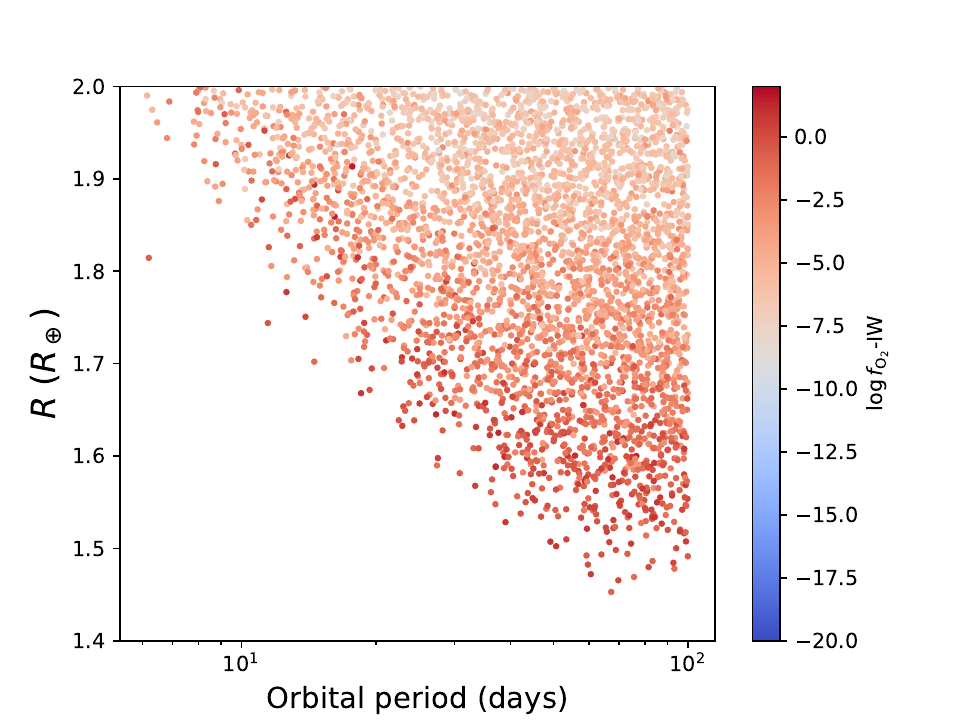}
\end{center}
\vspace{-0.1in}
\caption{Outgassing calculations informed by data-driven radius valley constraints.  Left column: reduced conditions corresponding to $\log{f_{\rm O_2}}$ and $\log{f_{\rm S_2}}$ being uniformly sampled between $-11$ to $-1$ (cgs units; corresponding to $10^{-17}$ to $10^{-7}$ bar). Right column: oxidized conditions corresponding to $\log{f_{\rm O_2}}$ and $\log{f_{\rm S_2}}$ being uniformly sampled between $-1$ to $2$ (cgs units; corresponding to $10^{-7}$ to $10^{-4}$ bar).  The top row shows the distributions of mean molecular weight, while the bottom row shows the corresponding oxygen fugacities relative to the iron-w\"{u}stite (IW) buffer (which is temperature- and pressure-dependent).  The elemental abundance of carbon is assumed to be solar (see text for more details).}
\vspace{0.2in}
\label{fig:population}
\end{figure*}

\subsection{Representative case studies of hybrid atmospheres}

Sub-Neptunes are exoplanets where the influence of the atmosphere and core are comparable, implying that they occupy equal volumes even if they have markedly different masses \citep{owen19}.  A hydrogen-dominated atmosphere has the same volume as a rocky core if it has a mass $\sim 1\%$ of the core mass \citep{owen19}.  The typical surface pressure of a sub-Neptune with a $4M_\oplus$ core and an atmospheric mass fraction of 1\% is
\begin{equation}
P = \frac{0.01 G M^2_{\rm core}}{4 \pi R_{\rm core}^4} \approx 17 \mbox{ kbar} ~\left( \frac{M_{\rm core}}{4 M_\oplus} \right)^2 \left( \frac{R_{\rm core}}{1.8 R_\oplus} \right)^{-4},
\end{equation}
where $G$ is Newton's gravitational constant.  We note that $10 \mbox{ kbar} = 1 \mbox{ GPa}$.  While seemingly large, such pressures are still much lower than what are required to render hydrogen metallic ($\sim 100$ GPa).

Figure \ref{fig:examples} shows examples of outgassing calculations for hypothetical exoplanets with $P_{\rm H_2} = 10$ kbar and 1 bar, corresponding to sub-Neptune-like and Earth-like surface pressures, respectively.  For illustration, we have assumed $f_{\rm O_2} = f_{\rm S_2} = 10^{-1}$ dyne cm$^{-2} = 10^{-7}$ bar.  

The dominant carbon carrier is methane for sub-Neptune-like surface pressures and carbon monoxide for Earth-like surface pressures.  Carbon dioxide is sub-dominant compared to methane for the model sub-Neptune, but this trend is reversed for the model super Earth.

At sub-Neptune-like surface pressures, the dominant silicon carrier is silane.  At Earth-like surface pressures, this switches to silicon monoxide.  This qualitative behavior occurs because the net reaction in equation (\ref{eq:silicon}) has 4 reactants and 2 products, which favors the forward reaction as pressure increases (Le Chatelier's principle).  Thus, the relative abundance of silicon monoxide versus silane may potentially act as a pressure diagnostic.

Hydrogen sulfide is the dominant sulfur carrier with abundances that are typically intermediate between those of water and carbon dioxide.  Sulfur dioxide is sub-dominant over a broad range of melt temperatures (2000--6000 K), confirming that it is not produced in abundance under the condition of chemical equilibrium \citep{tsai23}.

At sub-Neptune-like surface pressures, assuming ideal-gas behavior (corresponding to fugacity coefficients of unity) may lead to order-of-magnitude errors for the abundances of several molecules, especially carbon monoxide and carbon dioxide.  Nevertheless, the influence on the mean molecular weight is small, particularly when considering the trends.  Curiously, ignoring the solubility of water and molecular hydrogen in melt produces negligible differences in their abundances, but reduces the abundances of carbon-bearing molecules by a factor of several.  A fixed partial pressure of molecular hydrogen implies a smaller \textit{total} hydrogen budget when H$_2$ solubility is ignored.  For a fixed value of C/H, this then implies that the carbon budget is also smaller, which leads to lower abundances of CO, CO$_2$ and CH$_4$.  Overall, we conclude that the consideration of fugacity coefficients has a larger impact on accurately modeling sub-Neptunian atmospheres than solubility laws---at least, without an explicit treatment of the melt chemistry.  At Earth-like surface pressures, non-ideal gas behavior and solubility are negligible effects.

There is a trend of the mean molecular weight increasing from $\mu \approx 2$ to $\mu \approx 16$ as the melt temperature decreases.  This is slightly more pronounced for Earth-like, compared to the sub-Neptune-like, surface pressures.  Nevertheless, the dominant effect is the diminished temperature, rather than the diminished pressure.  Physically, these two properties are closely related: as the core becomes less massive, it retains less of its hydrogen envelope due to weaker gravity.  This in turn allows the exoplanet to cool more quickly.

While the outgassing calculations in Figure \ref{fig:examples} are illustrative, more insight is gained by elucidating the $P_{\rm H_2}(T)$ relationship and using it to inform outgassing calculations for a synthetic population of exoplanets.

\begin{figure*}[!ht]
\begin{center}
\vspace{-0.2in} 
\includegraphics[width=\columnwidth]{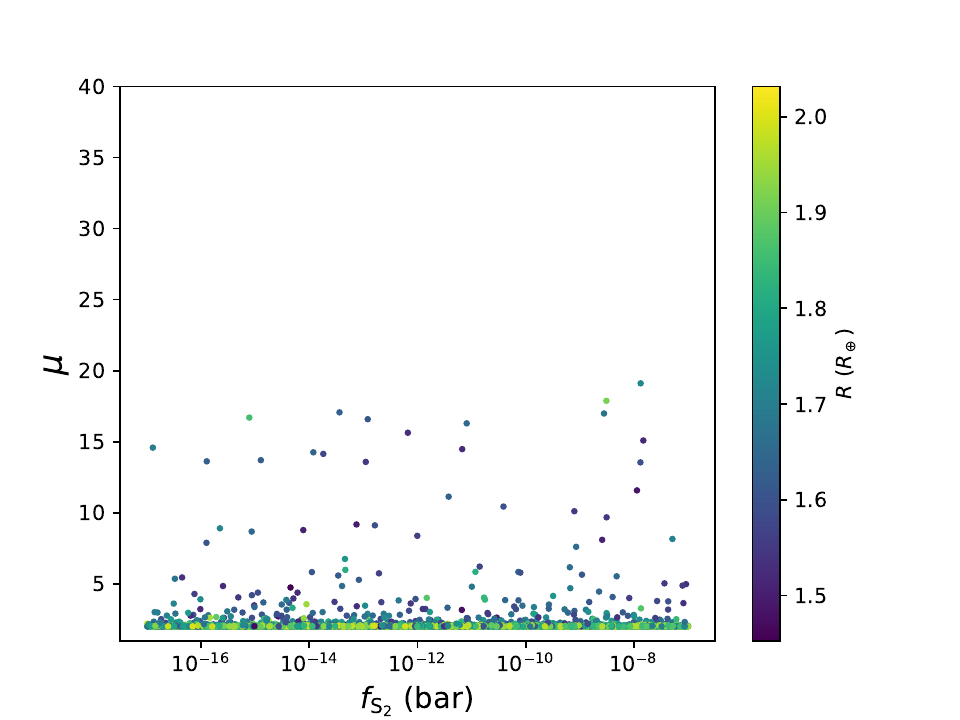}
\includegraphics[width=\columnwidth]{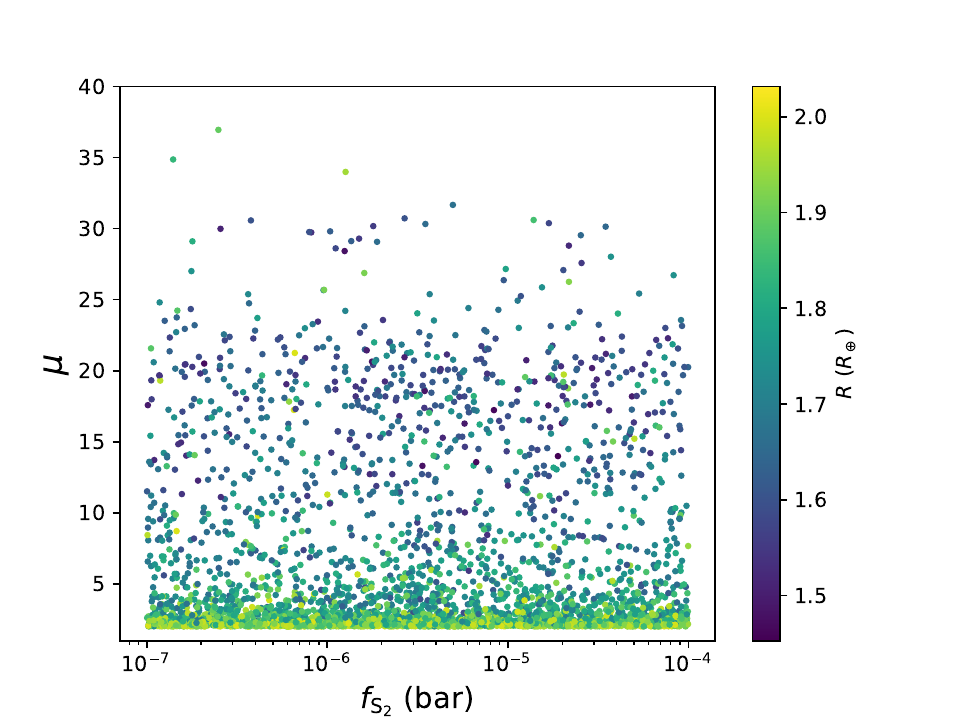}
\includegraphics[width=\columnwidth]{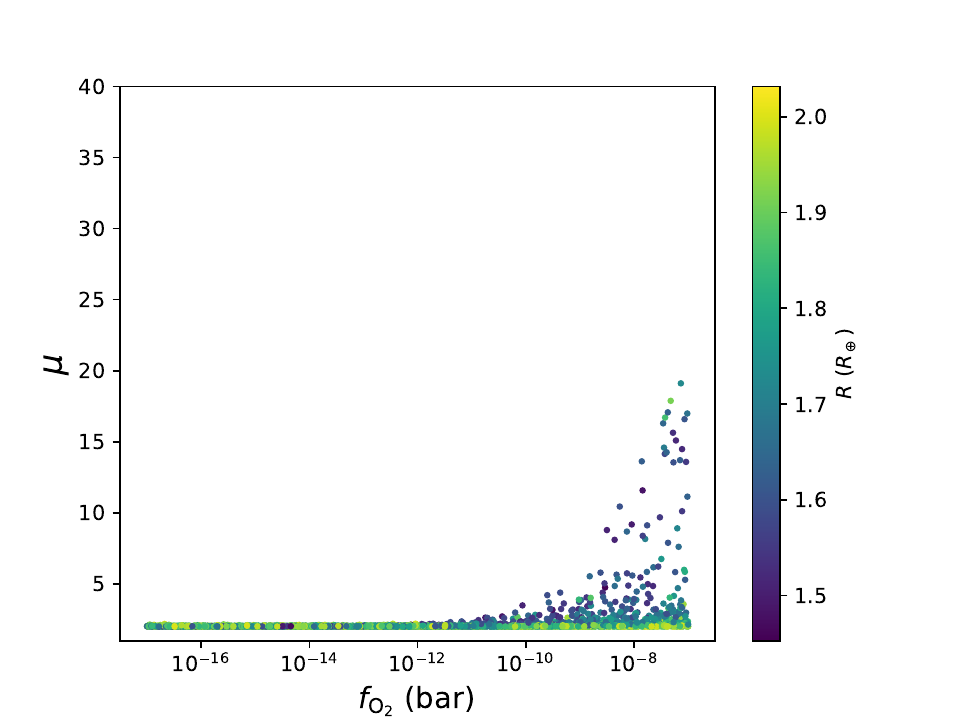}
\includegraphics[width=\columnwidth]{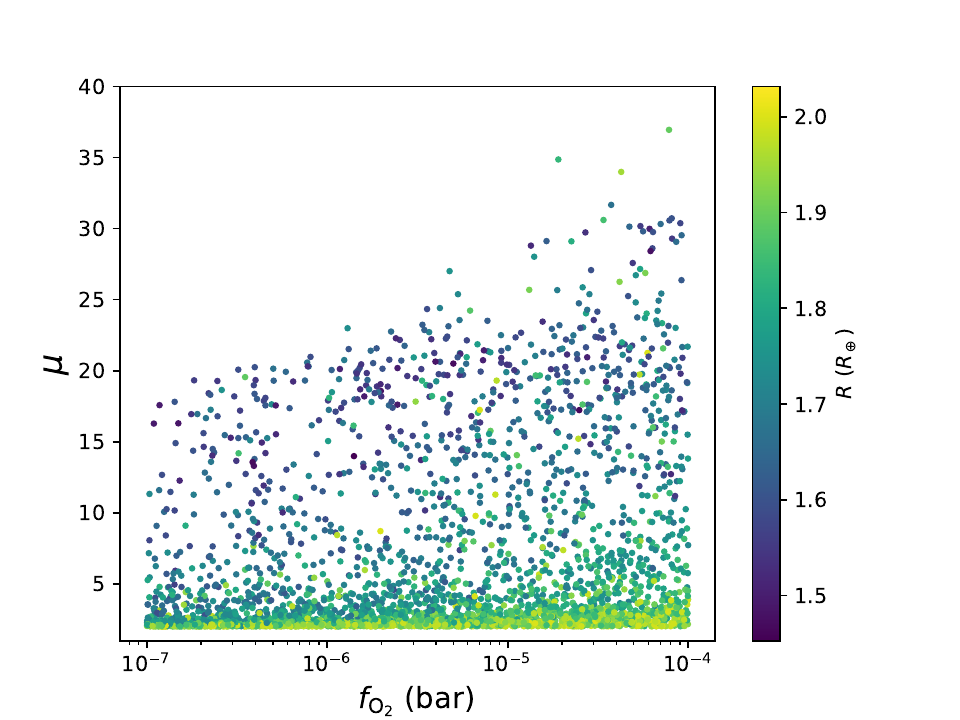}
\end{center}
\vspace{-0.1in}
\caption{Distributions of mean molecular weight as a function of the sulfur fugacity (top row) and oxygen fugacity (bottom row).  The left and right columns correspond to reduced and oxidized conditions, respectively (see text for more details).  These distributions are another way to visualise the calculations in Figure \ref{fig:population}.}
\vspace{0.2in}
\label{fig:population2}
\end{figure*}

\subsection{Population study of hybrid atmospheres}

We now use data-driven radius-valley constraints (Section \ref{subsect:radius_valley}) to compute outgassed chemistry for a population of exoplanets.   As larger exoplanets (the second peak of the radius distribution; $R \gtrsim 2.4 R_\oplus$) are expected to have hydrogen-dominated atmospheres, we curate the simulated sample to include only exoplanets with radii $R<2R_\oplus$ and $T> 2000$ K.  The former condition restricts the maximum surface pressure to be less than 77 kbar and the maximum melt temperature to be less than 6200 K.  The outcomes are independent of this curation as higher values of pressure and temperature correspond to canonical hydrogen-dominated atmospheres with $\mu \approx 2$.  The latter condition is the temperature threshold above which silicate species are relevant in the gaseous phase \citep{charnoz23,misener23}.

The melt mass fraction is given by equation (\ref{eq:chi_melt}).  As shown in Figure \ref{fig:Mratio}, $M_{\rm core}/M_{\rm atm}$ is informed by the calculations of \cite{ro21}.  In the absence of a first-principles theory on how well-mixed the rocky core is, we uniformly and randomly sample $M_{\rm melt}/M_{\rm core}$ between 0 and 1.

In the top row of Figure \ref{fig:population}, we show calculations of the mean molecular weight across radius and orbital period.  For what we term ``reduced conditions", $\log{f_{\rm O_2}}$ and $\log{f_{\rm S_2}}$ are uniformly sampled between $-11$ and $-1$ (cgs units; corresponding to $10^{-17}$ to $10^{-7}$ bar).  When they are sampled uniformly between $-1$ to $2$ (cgs units; corresponding to $10^{-7}$ to $10^{-4}$ bar), ``oxidized conditions" prevail.  These ranges of values are arbitrary and only serve to illustrate that the gradient of $\mu$ across radius and orbital period is robustly present.  We note that random sampling of parameters to mitigate incomplete theoretical understanding has been successfully applied to the modeling of debris disks \citep{ht10} and polluted white dwarfs \citep{vh20}.

In the bottom row of Figure \ref{fig:population}, we show the corresponding values of the sampled oxygen fugacities relative to the iron-w\"{u}stite buffer ($\log{f_{\rm O_2}}-\mbox{IW}$).  To compute the temperature and pressure dependence of IW, we use the empirical relation of \cite{hirschmann21}, which is strictly valid only up to 3000 K.  In other words, extrapolation beyond this temperature limit is sometimes employed.  The bottom-left panel of Figure \ref{fig:population} shows a qualitative trend that matches Figure 2c of \cite{cherubim25}, who model an oxidation gradient across the radius valley by considering atmospheric escape.  Quantitatively, both our sampled $\log{f_{\rm O_2}}$ (cgs units) and $\log{f_{\rm O_2}}-\mbox{IW}$ values are broadly consistent with those obtained by \cite{cherubim25} in their Figure 2c.  Further comparison is unwarranted as the models of \cite{cherubim25} make multiple assumptions about atmospheric escape and planet formation history, including assuming an initial atmospheric abundance of oxygen that is solar and a core mass fraction of 0.3.

\begin{figure*}[!ht]
\begin{center}
\vspace{-0.1in} 
\includegraphics[width=\columnwidth]{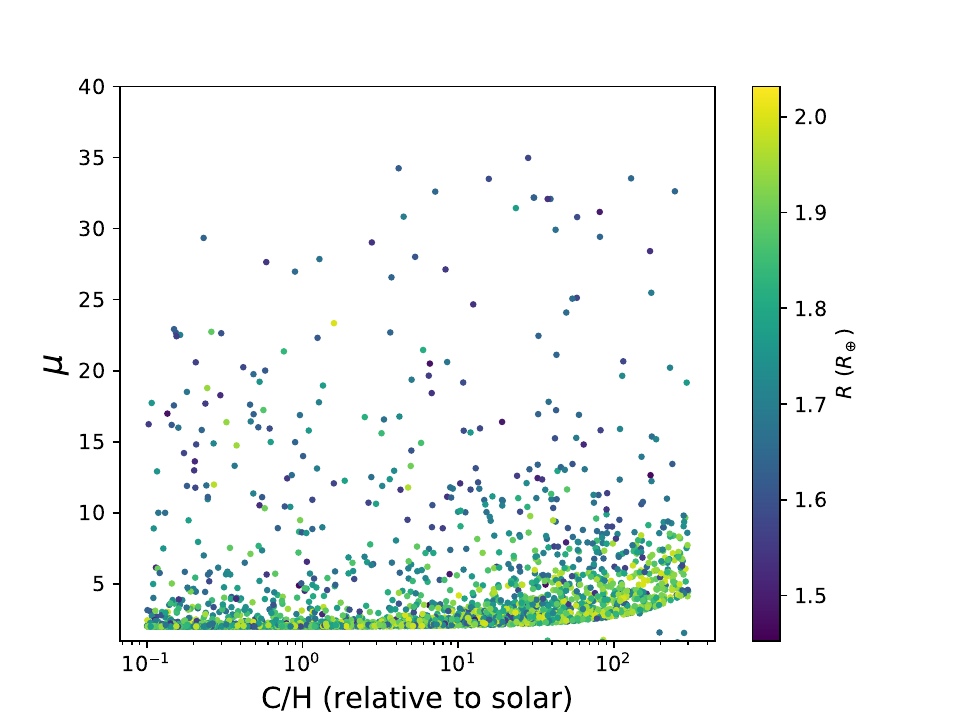}
\includegraphics[width=\columnwidth]{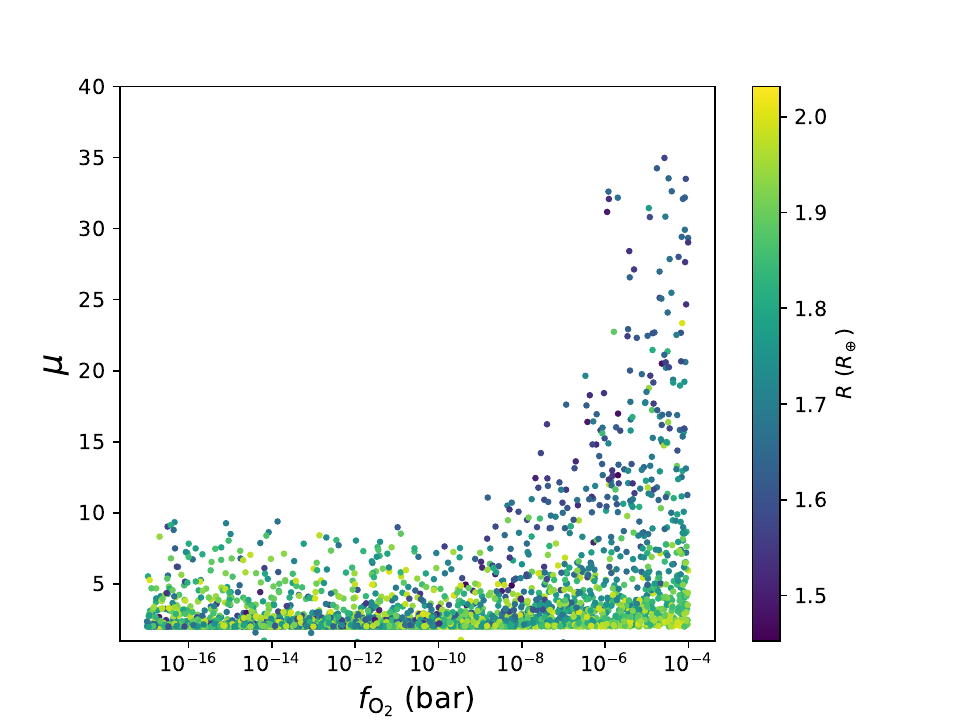}
\end{center}
\vspace{-0.1in}
\caption{Distributions of mean molecular weight as a function of the elemental abundance of carbon (C/H; left panel) and oxygen fugacity (right panel).}
\vspace{0.2in}
\label{fig:C_H}
\end{figure*}

Figure \ref{fig:population2} visualizes the computed mean molecular weights in a different way by plotting them versus $f_{\rm S_2}$ and $f_{\rm O_2}$.  Firstly, we see that the majority of exoplanets with $R \approx 2 R_\oplus$ have canonical hydrogen-dominated atmospheres.  This finding validates our curation of the simulated exoplanet sample.  Secondly, the upper envelope of $\mu$ values has no correlation with the value of the sulfur fugacity, regardless of whether reduced or oxidized conditions are assumed.  Thirdly, and by contrast, the upper envelope of $\mu$ values is sensitive to the oxygen fugacity, especially under reduced conditions.  \textit{This important finding proves that, while the diminished melt temperatures of lower-mass cores are responsible for the gradient of $\mu$, the strength of this gradient is controlled by $f_{\rm O_2}$.}  

The computed mean molecular weights in Figures \ref{fig:population} and \ref{fig:population2} are insensitive to the carbon enrichment of the hydrogen envelope (sometimes termed the ``metallicity"), which is assumed to be solar ($\mbox{C/H}=2.5 \times 10^{-4}$).  To prove this insensitivity, we execute an additional set of calculations where we allow $\log{\mbox{C/H}}$ to be uniformly sampled such that C/H varies from $0.1\times$ to $300\times$ its solar value.  Figure \ref{fig:C_H} shows that there is no correlation between the upper envelope of $\mu$-values and C/H.  However, when C/H is greater than about $20\times$ the solar value, it imposes a \textit{minimum} value of $\mu$---exactly as one would expect for a ``metal-rich" atmosphere.  Despite sampling such a broad range of C/H values, the correlation between $\mu$ and $f_{\rm O_2}$ remains qualitatively identical: the upper envelope of mean molecular weight values depends on the oxygen fugacity.  \textit{In other words, neither the gradient of $\mu$ nor its strength is primarily driven by the elemental abundance of carbon of the atmosphere.}  However, the \textit{range} of $\mu$ values may be extended by high values of C/H.  This is clearly seen when comparing the bottom left panel of Figure \ref{fig:population2} with the right panel of Figure \ref{fig:C_H}: for the former, the threshold value of the oxygen fugacity, for producing $\mu \gtrsim 2$ is $f_{\rm O_2} \gtrsim 10^{-11}$ bar; for the latter, there is no such threshold over the entire range of $f_{\rm O_2}$ sampled.

Figure \ref{fig:C_H} comes with two caveats.  Firstly, about $8\%$ of the outgassing calculations failed to attain numerical convergence when C/H is $\gtrsim 10\times$ solar, presumably because the initial guesses for the volume mixing ratios of molecules are inadequate (see Appendix \ref{append:CHOSSi}).  This implies that some of the empty regions in the $\mu$ versus C/H plot may actually be populated if all of the numerical calculations had converged.  Secondly, carbon-rich atmospheres may precipitate graphite \citep{moses13}, a process we do not include in our CHOSSi chemical system.  Graphite formation would remove carbon from the gaseous phase and reduce the mean molecular weight of the atmosphere.

\begin{figure*}[!ht]
\begin{center}
\vspace{-0.2in} 
\includegraphics[width=\columnwidth]{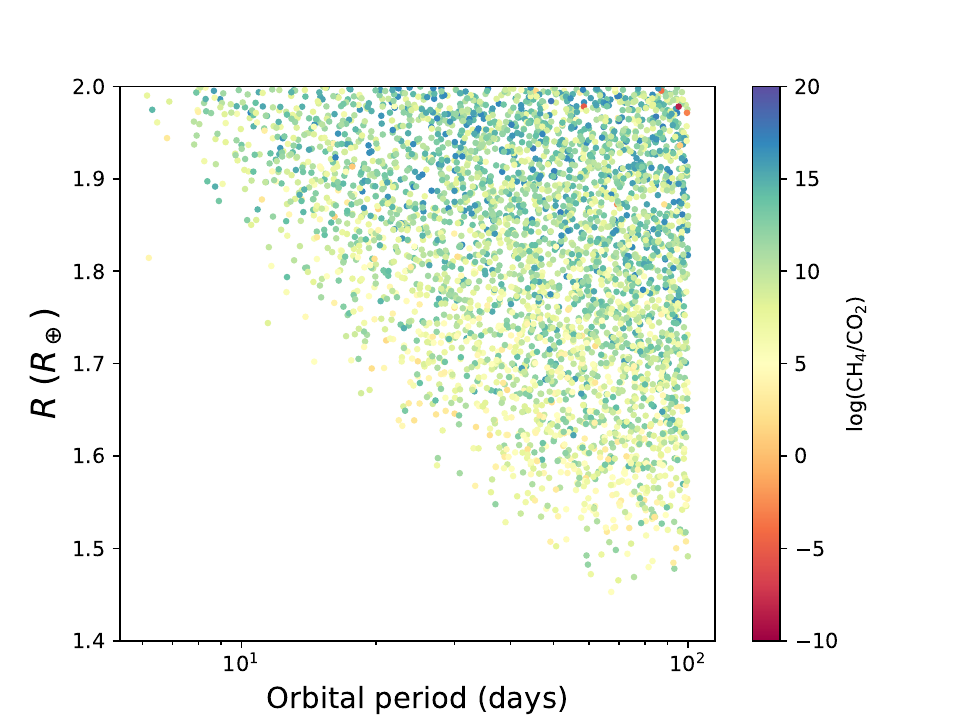}
\includegraphics[width=\columnwidth]{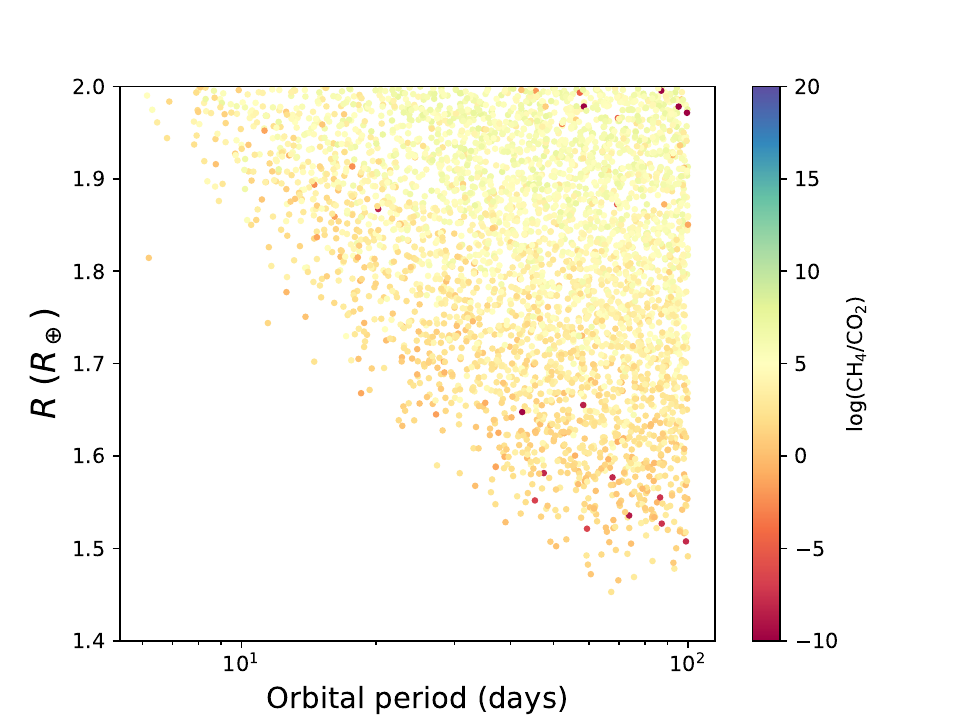}
\includegraphics[width=\columnwidth]{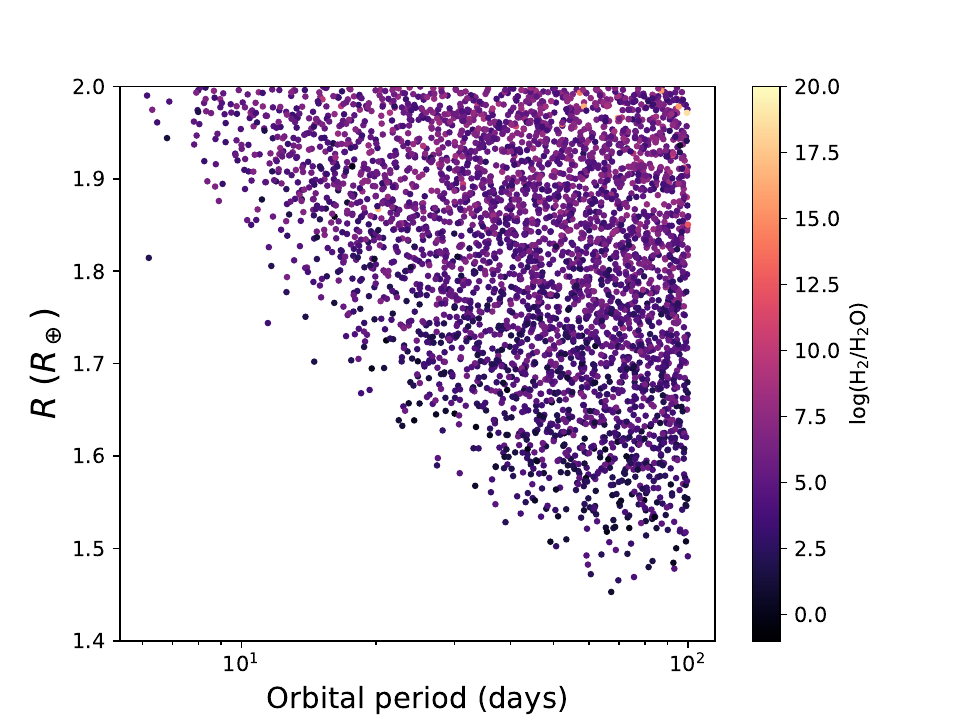}
\includegraphics[width=\columnwidth]{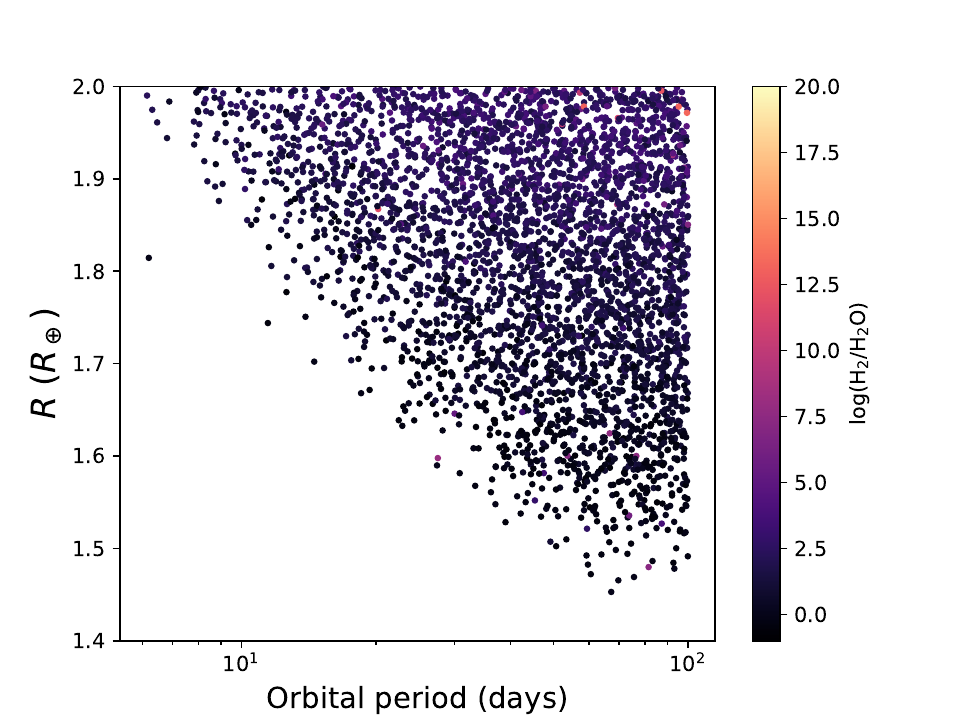}
\end{center}
\vspace{-0.1in}
\caption{Ratios of molecular abundances (by number) corresponding to the calculations in Figure \ref{fig:population}, where the left and right columns correspond to reduced and oxidized conditions, respectively (see text for more details).  The top and bottom row are for the ratios of methane (CH$_4$) to carbon dioxide (CO$_2$) and molecular hydrogen (H$_2$) to water (H$_2$O), respectively.}
\vspace{0.2in}
\label{fig:population_observables}
\end{figure*}

Figure \ref{fig:population_observables} shows the ratios of molecular abundances (by number) corresponding to the calculations in Figure \ref{fig:population}.  As expected, $X_{\rm CH_4}/X_{\rm CO_2}$ is highly sensitive to the oxidation state, which makes it a good diagnostic for inferring the oxygen fugacity.  In particular, it is possible to produce $X_{\rm CH_4}/X_{\rm CO_2} \sim 1$.  By contrast, $X_{\rm H_2}/X_{\rm H_2O}$ is less sensitive to the oxygen fugacity and fairly uniform across radius and orbital period.

\section{Discussion}
\label{sect:discussion}

\subsection{Summary}

The key conclusion of our current study is that geochemical outgassing from the molten rocky cores of sub-Neptunes naturally produces a diversity of mean molecular weights.  Sub-Neptunes with less massive cores retain less of their primordial hydrogen envelopes (against photo-evaporation driven by stellar irradiation), which lead to less retention of heat and lower melt temperatures at the surface of these cores.  The diminished surface pressures and melt temperatures robustly produce a gradient of mean molecular weight---simply due to thermodynamics and Gibbs free energies.  In other words, it is the temperature-pressure conditions alone that account for the existence of a gradient in $\mu$.  The \textit{strength} of this gradient is primarily controlled by the oxygen fugacity of the molten core.  To lowest order, the carbon enrichment of the hydrogen envelope plays no role in establishing the gradient of mean molecular weight, across the radius valley, or its strength.

\subsection{Comparison to previous work}

In the context of exoplanets, \cite{hl16} and \cite{ht16} previously derived analytical solutions for gaseous CHO and CHON chemical systems, respectively, but did not consider solubilities or non-ideal-gas behavior.  These systems of equations were designed for hot Jovian atmospheres and did not explicitly consider the oxygen fugacity as a parameter.  By contrast, \cite{th24} considered a CHONS chemical system from the perspective of outgassed atmospheres.  This study included activities and fugacities, but did not consider solubilities and mass budgets for hydrogen, carbon, silicon and oxygen.

The work of \cite{misener23} provides the closest comparison to the current study, as they solved for a hydrogen-oxygen-silicon (HOSi) system in chemical equilibrium under ideal-gas conditions and ignoring solubilities.  When calculating the abundance of silane, it is crucial to consider the solubility of water in melt as its suppression allows for silane to exist at $\sim 10\%$ abundance \citep{ito25}.  In solving for the thermal structure of sub-Neptunes, \cite{misener23} claimed that the decreasing abundance of silicon monoxide, as the temperature decreases towards higher altitudes, stabilises the atmosphere against convection. The transport of heat is assumed to occur via radiation and conduction, which these authors model by defining an effective opacity that takes the harmonic mean of the opacities, from both processes, in the diffusive limit.

This approach of modeling heat transport is in contrast to that of \cite{yu21}, where the atmosphere is visualised as having three regions: a deep component in chemical equilibrium, an intermediate component dominated by atmospheric mixing (parametrized by a diffusion coefficient in one dimension) and an upper atmosphere driven by photochemistry.  Atmospheric mixing includes large-scale circulation, which occurs even in the absence of convection due to latitudinal entropy gradients driven by stellar heating \citep{heng11}.  If the dynamical timescale is less than the chemical timescale, then the molecular abundances set by the deep atmosphere (in chemical equilibrium) are quenched \citep{pb77,smith98,vm11} and transported to the intermediate and upper atmospheres \citep{yu21}.  Therefore, chemical equilibrium may not hold throughout the entire atmosphere, as \cite{misener23} have assumed.  It is unclear how to estimate the ``quench point" \citep{vm11,tsai17} in the model atmospheres of \cite{misener23}, which is the pressure/altitude at which the dynamical and chemical timescales are equal.  Different chemical species have different quench points and thus how their abundances are propagated from the deep atmosphere to the intermediate and upper atmospheres differ \citep{moses11,moses13,tsai17}.  At pressures lower than that of the quench point, chemical equilibrium is a poor assumption.  A treatment beyond chemical equilibrium in the atmosphere is necessary for predicting the observable atmospheric composition.

\subsection{What about miscible cores?}

\cite{benneke24} have previously suggested that the enhanced value of the mean molecular weight inferred in the sub-Neptune TOI-270d may be attributed to it having a ``miscible core".  While matter at pressures $\sim \mbox{GPa}$ certainly exists in a supercritical state, the main effect driving the enhanced mean molecular weight is the geochemical outgassing of these molten cores, from melts with temperatures of about 2000-4000 K, into the primordial hydrogen envelope.  This conclusion is robust as it only depends on thermodynamics and Gibbs free energies, and occurs over a broad range of oxygen fugacities of the core. 

\subsection{Limitations and opportunities for future work}

\subsubsection{Activities, fugacities and solubilities}

Geochemical outgassing calculations of super Earths and sub-Neptunes are limited mostly by the scarcity or non-existence of thermodynamic quantities needed to quantify non-ideal-gas behavior (fugacities), non-ideal mixing of gaseous components (activities) and the tendency for certain gaseous species to dissolve in melt (solubilities).  While fugacities are not necessary for super Earths with surface pressures below $\sim 1$ kbar, they are needed for sub-Neptune surface pressures $\sim 10$ kbar.  

In the current study, we have not considered activity coefficients at all, as they depend not only on temperature and pressure but also on the composition of the gaseous mixture.  Activity coefficients do not exist for CHOSSi systems.  Of particular concern is that hydrogen sulfide may be highly soluble in melt, but there currently exists no experimental data to formulate a reliable solubility law for H$_2$S.  A  key limitation of the current study is the use of fugacity coefficients and solubility laws outside their intended ranges of temperatures (i.e., extrapolation).  

All of these avenues are ripe for future work---both experimental and theoretical.  When some of these temperature regimes are inaccessible to experiment, simulations of molecular dynamics can fill in this methodological gap (e.g., \citealt{dufils20}).  

\subsubsection{Outgassing model}

Despite its spectral importance for sub-Neptunes \citep{hu21,tsai21,yu21}, we have not included nitrogen because there is no clear way to specify the primordial abundance of nitrogen across the distributions of radius and orbital period of small exoplanets.  Including nitrogen will not qualitatively alter the key conclusions of the current study.  Another obvious avenue for future work is to develop an explicit treatment of the melt chemistry that is appropriate for the temperature-pressure conditions of sub-Neptunian cores (see discussion in Section 4.1 of \citealt{ito25}).

\subsubsection{Generalizing \cite{ro21}}

\cite{ro21} derived data-driven properties of the underlying small exoplanet population by assuming hydrogen-dominated envelopes with solar metallicity.  This is a good assumption for exoplanets associated with the second peak ($R \approx 2.4 R_\oplus$) of the radius distribution, but starts to break down for smaller exoplanets across the radius valley.  For atmospheres with $\mu \gg 2$, the geochemical outgassing and radius valley calculations are not consistent with each other.  With the theoretical framework introduced in the current study, outgassing may be incorporated into the calculations of \cite{ro21} to allow for the consideration of hybrid atmospheres \citep{th24}.

\subsubsection{Large-scale transport in sub-Neptunian atmospheres}
\label{subsect:mixing}

Without more detailed investigations, it is not obvious whether the thick atmosphere of a sub-Neptune is well-mixed.  While the hydrogen-dominated atmosphere and rocky core occupy comparable volumes \citep{owen19}, the core dominates the mass budget.  If a substantial fraction of the core engages in geochemical outgassing, then its influence on the atmosphere motivates a deeper investigation.

The current study predicts the lower boundary condition for atmospheric chemistry.  Despite the suggestion by \cite{misener23} that convection near the rocky core of a sub-Neptune is suppressed because of compositional gradients, this needs to be investigated more thoroughly because the chemical and dynamical timescales may be comparable (and thus chemical equilibrium is a poor assumption).  The sub-Neptunes in our curated synthetic sample have equilibrium temperatures between about 260 and 1200 K.  Some of these equilibrium temperatures---and therefore instellations---are high enough that large-scale circulation induced by stellar heating may penetrate deeply into the atmosphere.  Atmospheric circulation is essentially driven by entropy (or potential temperature) gradients between the equator and poles of an irradiated sub-Neptune \citep{heng11}---whether it penetrates down to high enough pressures to interact with deep convection near the core is unknown.  

Simulating such ``mixed" dynamics, deep within a sub-Neptune, is particularly challenging as it requires prohibitive numerical integration times for the deep atmosphere to ``spin up" and reach a statistical steady state \citep{sm19}.  Such simulations are needed to understand the interplay---if any---between convection near the core and large-scale circulation that encompasses the photosphere of a sub-Neptune.

Ultraviolet radiation from the star additionally modifies the observed atmospheric chemistry (photochemistry) of a sub-Neptune \citep{hu21,tsai21,yu21}.  While photochemistry alters the partitioning of carbon, hydrogen, oxygen, nitrogen, sulfur, silicon, etc, among the different molecular species, it is unlikely that it will reduce the mean molecular weight by an order of magnitude.  Nevertheless, this intuition needs to be verified by photochemical calculations.

Generally, the dynamical, chemical and radiative timescales of the atmosphere of a sub-Neptune may be comparable, implying that self-consistent, coupled simulations of dynamics, radiative transfer and photochemical kinetics may be necessary to fully understand the precise relationship between the outgassed and photospheric chemistry.

\vspace{0.1in}
{\scriptsize KH developed the theoretical formalism, derived the equations, performed the numerical calculations, made the figures and led the writing of the manuscript.  MT co-developed the theoretical formalism, checked the mathematical derivations of KH and formulated both the equations and the computer code for calculating fugacity coefficients.  JEO provided guidance on radius valley constraints and unpublished calculations from \cite{ro21}.  KH thanks Fabrice Gaillard for clarifying the solubility law of molecular hydrogen published in \cite{gaillard22}, Janine Birnbaum for pointing out the tension between Dalton's law and Newton's second law, Nicolas Sartor for suggesting references on solubility laws, Steve Mojzsis for stimulating conversations and an anonymous referee for stimulating comments that improved the quality of the manuscript.  KH acknowledges partial financial support from the European Research Council (ERC) Geoastronomy Synergy Grant (grant number 101166936; PIs: Gaillard, Heng and Mojzsis).}

\appendix
\section{Gibbs free energies of formation}
\label{append:gibbs}

We interpret $\Delta G$ as the Gibbs free energy of formation (see Section 3.1 of \citealt{th24} for a discussion).  The JANAF database\footnote{\texttt{https://janaf.nist.gov}} tabulates Gibbs free energies of formation for various chemical species at a reference pressure of $P_0=1$.  As \textit{molar} Gibbs free energies are provided (with physical units of kJ mol$^{-1}$), we choose to use the universal gas constant of ${\cal R} = 8.3144621$ J K$^{-1}$ mol$^{-1}$ in our calculations.  Section 2.3 of \cite{hl16} provides a detailed discussion of how to treat the physical units carefully.  JANAF lists the Gibbs free energy of formation for H$_2$ and O$_2$ to be zero; for S$_2$, it is zero only for $T \ge 900$ K.  

\begin{figure}[!ht]
\begin{center}
\vspace{-0.11in}
\includegraphics[width=0.49\columnwidth]{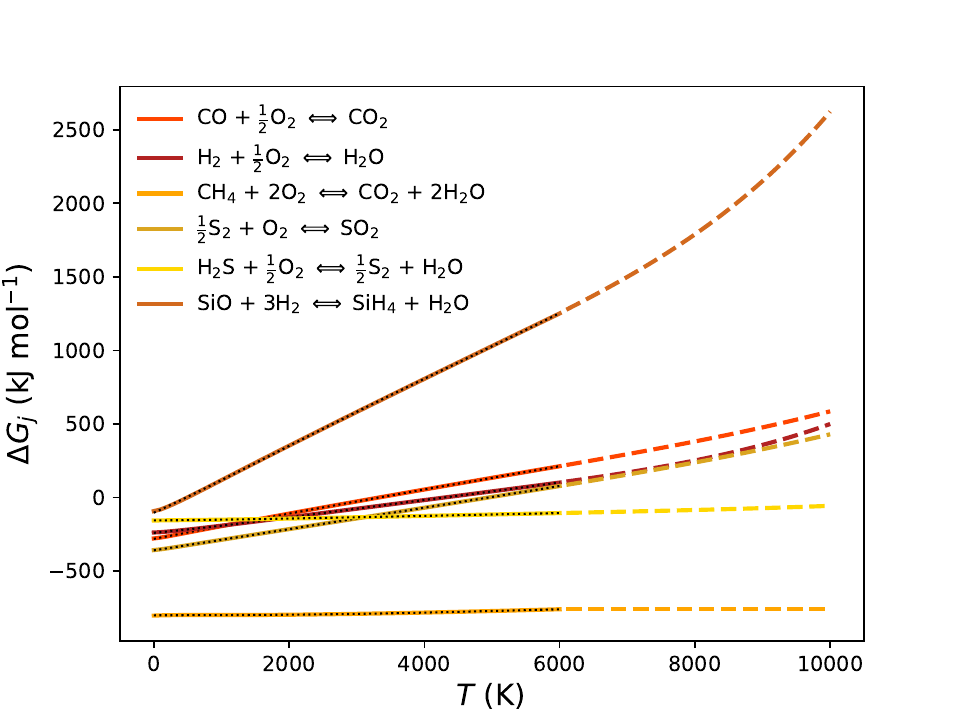}
\includegraphics[width=0.49\columnwidth]{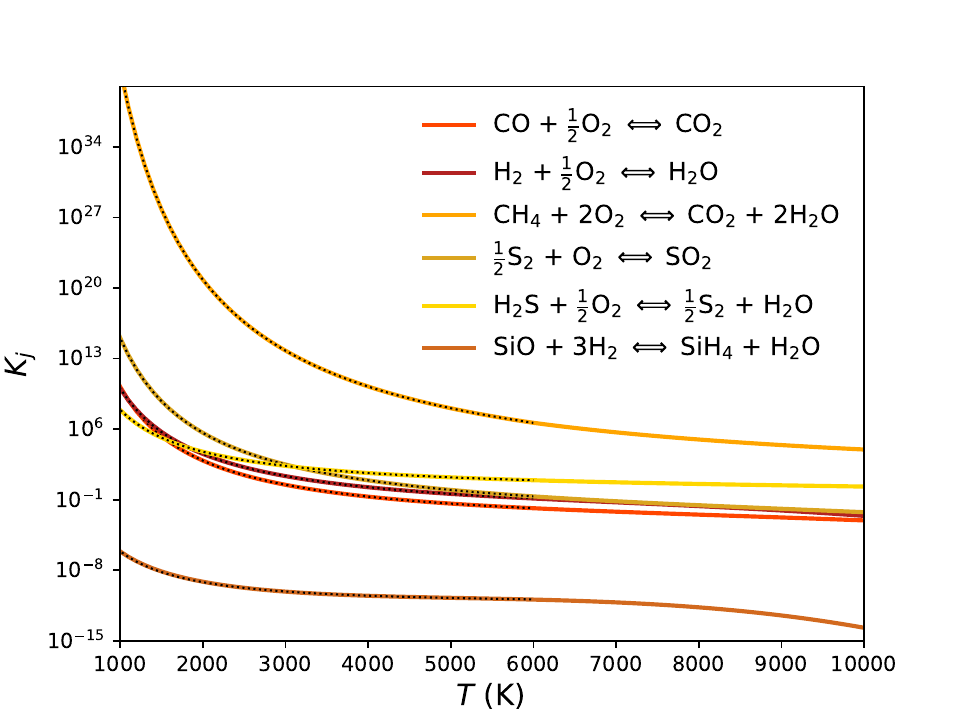}
\end{center}
\vspace{-0.1in}
\caption{Left panel: Gibbs free energies of formation for the 6 net chemical reactions stated in equations (\ref{eq:CHOS}) and (\ref{eq:silicon}) as constructed using the JANAF database.  The dotted curves are fits performed using Chebyshev polynomials (see text for details) for $0 \le T \le 6000$ K.  The dashed curves are the same fits extrapolated to $6100 \le T \le 10^4$ K, except for $\Delta G_3$ where we assume its value to be constant (and equal to the value at 6000 K).  Right panel: equilibrium constants constructed using the Chebyshev polynomial fits, where the curves beyond 6000 K are extrapolations.  The dotted curves are the equilibrium constants constructed using the original JANAF data.}
\vspace{-0.05in}
\label{fig:gibbs}
\end{figure}

Figure \ref{fig:gibbs} shows $\Delta G_j$ for $j=1$--$6$ for the 6 net chemical reactions stated in equations (\ref{eq:CHOS}) and (\ref{eq:silicon}).  For convenience, we fit these $\Delta G_j$ curves using Chebyshev polynomials (denoted by ${\cal T}_k$ for a Chebychev polynomial of the $k$-th order),
\begin{equation}
\Delta G_j = \sum_{k=0}^4 {\cal C}_k ~{\cal T}_k(T).
\end{equation}
The fits are performed using the \texttt{polynomial.chebyshev.chebfit} function in the Python \texttt{numpy} programming package.  For scientific reproducibility, Table \ref{tab:gibbs} reports the fit coefficients ${\cal C}_k$.

As discussed in the main text, sub-Neptunes may have temperatures at the atmosphere-core interface reaching $\sim 10^4$ K, whereas the JANAF database provides Gibbs free energies only up to 6000 K.  Since the curves of $\Delta G_j$ are smooth and well-behaved (Figure \ref{fig:gibbs}), we use the fitting functions to extrapolate them for $T>6000$ K except for $\Delta G_3$ as the resulting extrapolated function is non-monotonic.  In this case, we simply use the value of $\Delta G_3$ at 6000 K for $T>6000$ K.  Figure \ref{fig:gibbs} also shows the equilibrium constants $K_j$ constructed using equation (\ref{eq:keq}), where the extrapolated portions of these curves appear reasonable by visual inspection.

\begin{table}
\label{tab:gibbs}
\begin{center}
\caption{Coefficients of Chebyshev polynomial fits}
\begin{tabular}{lccccc}
\hline
\hline
Reaction & ${\cal C}_0$ & ${\cal C}_1$ & ${\cal C}_2$ & ${\cal C}_3$ & ${\cal C}_4$ \\
\hline
1 & $-2.81990035 \times 10^{2}$ & $8.63857116 \times 10^{-2}$ & $1.14004268 \times 10^{-7}$ & $-8.64090135 \times 10^{-11}$ & $4.07761984 \times 10^{-15}$ \\
2 & $-2.41212701 \times 10^{2}$ & $4.26423144 \times 10^{-2}$ & $3.75831753 \times 10^{-6}$ & $-3.67706178 \times 10^{-10}$ & $1.29138659 \times 10^{-14}$ \\
3 & $-8.02022252e \times 10^{2}$ & $1.54466780 \times 10^{-3}$ & $-2.87989051 \times 10^{-7}$ & $1.29417601 \times 10^{-10}$ & $-5.75526461 \times 10^{-15}$ \\
4 & $-3.60405567 \times 10^{2}$ & $7.06878078 \times 10^{-2}$ & $5.98577039 \times 10^{-7}$ & $-7.12203913 \times 10^{-11}$ & $3.09693938 \times 10^{-15}$ \\
5 & $-1.56860063 \times 10^{2}$ & $3.78126944 \times 10^{-3}$ & $8.70325866 \times 10^{-7}$ & $-5.98505785 \times 10^{-11}$ & $1.58640487 \times 10^{-15}$ \\
6 & $-1.04479777 \times 10^2$ &  $2.10409072 \times 10^{-1}$ & $7.36113799 \times 10^{-6}$ & $-9.39597816 \times 10^{-10}$ & $3.63793942 \times 10^{-14}$ \\
\hline
\hline
\end{tabular}\\
\vspace{0.05in}
Note: reactions 1 to 6 are ordered according to equations (\ref{eq:CHOS}) and (\ref{eq:silicon}).  \\ The Chebyshev polynomials take as input the temperature (with physical units of K) and $\Delta G_j$ is fitted in physical units of kJ mol$^{-1}$.
\end{center}
\end{table}

\section{Fugacity coefficients for H$_2$, H$_2$O, CO, CO$_2$ and CH$_4$}
\label{append:fugacity}

The basic equations and fitting functions for computing the fugacity coefficients are provided in this appendix.  They are valid for $P=1$ bar to $50$ kbar and $373 \le T \le 1873$ K \citep{hp91}. 

\subsection{Basic equations}
\label{sect:general}

For a pure gas $i$, its molar Gibbs free energy\footnote{With physical units of kJ mol$^{-1}$.} (or chemical potential), is (cf. pg. 185 of \citealt{devoe})
\begin{equation} \label{eq: general_G}
    G_i(P, T) = G_i(P_0, T) + \int_{P_0}^{P}V ~dP,
\end{equation}
where $V$ is the molar volume\footnote{In the fitting formulae described in this appendix, $V$ is computed with physical units of kJ (kbar)$^{-1}$ mol$^{-1}$ = 10 cm$^3$ mol$^{-1}$.}.  If the gas behaves ideally, we have $PV = {\cal R} T$ where ${\cal R} = 8.314472 \times 10^{-3}$ kJ K$^{-1}$ mol$^{-1}$ is the universal gas constant.  Evaluating the integral for an ideal gas, we obtain the familiar expression,
\begin{equation} \label{eq: ideal_G}
    G_i(P, T) = G_i(P_0, T) + {\cal R} T ~\ln{\left(\frac{P}{P_0}\right)}.
\end{equation}

If the gas does not behave ideally, then equation \eqref{eq: ideal_G} does not hold.  However, one can always \textit{force} the functional form,
\begin{equation} \label{eq: real_G}
    G_i(P, T) = G_i(P_0, T) + {\cal R} T\ln{\left(\frac{f}{P_0}\right)},
\end{equation}
equate it to equation \eqref{eq: general_G} and compute the fugacity,
\begin{equation}
f = P_0 ~\exp{\left[\frac{1}{{\cal R} T} \int^P_{P_0} V ~dP \right]}.
\end{equation}
As long as the functional form of the equation of state, $V=V(P,T)$, is known, the fugacity of a pure gas at any pressure and temperature may be computed via numerical integration (cf. pg. 186 of \citealt{devoe}).  Upon computing $f$, the fugacity coefficient may obtained via $\phi = f/P$.

In practice, it is often the fugacity coefficient that is directly computed via (\citealt{kite19}; pg. 187 of \citealt{devoe})
\begin{equation} \label{eq: phi_Z}
    \ln\phi = \int_0^P \frac{Z-1}{P} ~dP,
\end{equation}
where the compressibility factor is given by
\begin{equation} \label{eq: compress_factor}
    Z=\frac{PV}{{\cal R} T}.
\end{equation}
For ideal gases, we have $Z=1$ and thus $\phi=1$ by construction.  Non-ideal gases have $Z \ne 1$.  Effectively, $Z$ is the functional expression of non-ideal equations of state (EoS).

For the rest of this appendix, $P$ is expressed in physical units of kbar for all subsequent fitting formulae.  $V$ is expressed with a subscript that labels the acronym of the EoS being used.

\subsection{Specific equations of state}
\label{sect:specific}
\cite{rk49} proposed a non-ideal EoS that was subsequently improved to be the modified Redlich-Kwong (MRK) EoS \citep{hp91},
\begin{equation} \label{eq: mrk}
    P = \frac{{\cal R} T}{V_{\rm MRK}-b} - \frac{a}{V_{\rm MRK}(V_{\rm MRK}+b)\sqrt{T}},
\end{equation}
where $a$ and $b$ are coefficients calibrated on experimental data. 

In the current study, the non-ideal EoS used is a further improvement by compensating for the tendency of MRK to overestimate the molar volume under high pressures.  It is termed the ``COmpensated Redlich-Kwong" (CORK) EoS \citep{hp91, hp98}. CORK makes the following virial-type compensation \citep{hp91, hp98},
\begin{equation} \label{eq: cork}
    V_{\rm CORK} = V_{\rm MRK} + c(P-P^{\circ}) + d(P-P^{\circ})^{1/2} + e(P-P^{\circ})^{1/4},
\end{equation}
where $c$, $d$ and $e$ are empirical coefficients.  The quantity $P^{\circ}$ is the threshold pressure above which $V_{\rm CORK}$, instead of $V_{\rm MRK}$, is used.  The preceding equation is stated for completeness and is not used in the calculation of the fugacity coefficient.

Using the MRK EoS to calculate $Z$, \cite{rk49} provided the fugacity coefficient expression based on MRK, which was reformulated by \cite{hp91},
\begin{equation} \label{eq: phi_mrk}
    \ln \phi_{\rm MRK} = Z - 1.0 - \ln(Z - B) - A\ln\left(1+\frac{B}{Z} \right),
\end{equation}
where $B=bP/{\cal R}T$ and $A=a/B{\cal R}T^{3/2}$.

When $P>P^{\circ}$, equation (\ref{eq: mrk}) is used to compute $V_{\rm MRK}$\footnote{In the appendix of \cite{hp91}, it was hinted that $V_{\rm CORK}$, rather than $V_{\rm MRK}$, is used to determine $Z$ in equation (\ref{eq: phi_mrk}), but this is incorrect and inconsistent with their equations (A.2) and (A.3).}, which is re-expressed as $Z$ using equation (\ref{eq: compress_factor}).  With $Z$ in hand, equation (\ref{eq: phi_mrk}) allows one to compute $\phi_{\rm MRK}$.  The virial-type correction to the MRK-based fugacity coefficient is as follows \citep{hp91, hp98}:
\begin{equation} \label{eq: phi_cork}
    \ln \phi_{\rm CORK} = \ln \phi_{\rm MRK} + \ln \phi_{\rm virial},
\end{equation}
\begin{equation} \label{eq: phi_corr}
\begin{split}
\ln\phi_{\rm virial} &= \frac{1}{{\cal R}T}\left[\frac{c}{2}(P-P^{\circ})^2 + \frac{2d}{3}(P-P^{\circ})^{3/2} + \frac{4e}{5}(P-P^{\circ})^{5/4} \right].
\end{split}
\end{equation}
The CORK EoS provides the empirical coefficients $a$, $b$, $c$, $d$, $e$, and $P^{\circ}$, which will be described in the following.

\subsubsection{Empirical coefficients for EoS of H$_2$O}

The critical temperature of water is 695 K, below which gaseous and liquid H$_2$O are distinguishable. Below 695 K, \cite{hp91} devised two expressions for $a$.  However, since the temperatures we are exploring in the current study are well above 695 K, this is of no concern for us.  The expression for $a$ is
\begin{equation}
    a = \\
\begin{cases}
    &a_0+a_1(T_0-T)+a_2(T_0-T)^2 +a_3(T_0-T)^3, \text{ if } T \leq T_0, \\
    &a_0+a_4(T-T_0)+a_5(T-T_0)^2 +a_6(T-T_0)^3, \text{ if } T > T_0,
\end{cases}
\label{eq:a0}
\end{equation}
where $T_0 = 673$ K.  The empirical coefficients $b$, $c$, $d$, $e$ and $P^{\circ}$ for the CORK EoS of H$_2$O are tabulated in Table \ref{tab:h2o} (reproduced from Table 1 from \citealt{hp91}, which was updated by \citealt{hp98}).

\begin{table}
\begin{center}
\caption{Empirical coefficients for CORK EoS of H$_2$O}
\label{tab:h2o}
\begin{tabular}{lcrc}
\hline
\hline
    &  & Value & Unit    \\ 
\hline
$a$ &  &    & kJ$^2$ kbar$^{-1}$ K$^{1/2}$ mol$^{-2}$  \\
    & $a_0$ & 1113.4 & $\dagger$ \\
    & $a_1$ & $-0.88517$ & $\dagger$ \\
    & $a_2$ & $4.53\times10^{-3}$ & $\dagger$ \\
    & $a_3$ & $-1.3183\times10^{-5}$ & $\dagger$ \\
    & $a_4$ & $-0.22291$ & $\dagger$ \\
    & $a_5$ & $-3.8022\times10^{-4}$ & $\dagger$ \\
    & $a_6$ & $1.7791\times10^{-7}$ & $\dagger$ \\
$b$ &  &  1.465 & kJ kbar$^{-1}$ mol$^{-1}$\\
$c$ &  &  $1.9853\times10^{-3}$  & kJ kbar$^{-2}$ mol$^{-1}$\\
$d$ &  &  $-8.9090\times10^{-2}$ & kJ kbar$^{-3/2}$ mol$^{-1}$\\
$e$ &  &  $8.0331\times10^{-2}$ & kJ kbar$^{-5/4}$ mol$^{-1}$\\
$P^{\circ}$ &  &  2 & kbar\\
\hline
\hline
\end{tabular}\\
\vspace{0.1in}
$\dagger$: Physical units are omitted for convenience. The coefficients $a_j$ take on the appropriate units in equation (\ref{eq:a0}) such that $a$ has units of kJ$^2$ kbar$^{-1}$ K$^{1/2}$ mol$^{-2}$.
\end{center}
\vspace{-0.2in}
\end{table}

\subsubsection{Empirical coefficients for EoS of CO$_2$}
\label{subsect:EOS_CO2} 

The empirical coefficients for the CORK EoS of CO$_2$ are:
\begin{equation}
\begin{split}
a &= a_0 + a_1 T + a_2 T^2, \\
c &= c_0 + c_1 T, \\
d &= d_0 + d_1 T.
\end{split}
\label{eq:CO2_EOS_coeff}
\end{equation}
The coefficients $a_j$, $c_j$ and $d_j$ are stated in Table \ref{tab:co2} (reproduced from Table 1 of \citealt{hp91}, which was updated by \citealt{hp98}).

\begin{table}
\begin{center}
\caption{Empirical coefficients for CORK EoS of CO$_2$}
\label{tab:co2}
\begin{tabular}{lcrc}
\hline
\hline
    &  & Value & Unit    \\ 
\hline
$a$ &  &    & kJ$^2$ kbar$^{-1}$ K$^{1/2}$ mol$^{-2}$ \\
    & $a_0$ & 741.2 & $\dagger$ \\
    & $a_1$ & $-0.10891$ & $\dagger$ \\
    & $a_2$ & $-3.4203\times10^{-4}$ & $\dagger$ \\
$b$ &  &  3.057 & kJ kbar$^{-1}$ mol$^{-1}$\\
$c$ &  &    & kJ kbar$^{-2}$ mol$^{-1}$\\
    & $c_0$ & $5.40776\times10^{-3}$ & $\dagger$ \\
    & $c_1$ & $-1.59046\times10^{-6}$ & $\dagger$ \\
$d$ &  &    & kJ kbar$^{-3/2}$ mol$^{-1}$\\
    & $d_0$ & $-1.78198\times10^{-1}$ & $\dagger$ \\
    & $d_1$ & $2.45317\times10^{-5}$ & $\dagger$ \\
$e$ &  &  0         & kJ kbar$^{-5/4}$ mol$^{-1}$\\
$P^{\circ}$ &  &  5.0  & kbar\\
\hline
\hline
\end{tabular}\\
\vspace{0.1in}
$\dagger$: Physical units are omitted for convenience. The coefficients $a_j$, $c_j$ and $d_j$ take on the appropriate physical units when used to compute $a$, $b$, $c$ and $d$ in equation (\ref{eq:CO2_EOS_coeff}).
\end{center}
\end{table}

\subsubsection{Empirical coefficients for EoS of H$_2$, CO and CH$_4$}

As H$_2$, CO and CH$_4$ have low critical temperatures (Table \ref{tab:cp}), the empirical CORK EoS may be simplified \citep{hp91},
\begin{equation} \label{eq: approx_cork}
    V_{\rm CORK} = \frac{{\cal R} T}{P} + b - \frac{a{\cal R}\sqrt{T}}{({\cal R}T+bP)({\cal R}T+2bP)} + c\sqrt{P} + dP,
\end{equation}
which in turn simplifies the expression for the fugacity coefficient,
\begin{equation} \label{eq: approx_f}
    {\cal R} T \ln \phi = bP + \frac{a}{b\sqrt{T}} ~\ln{\left(\frac{{\cal R}T+bP}{{\cal R}T+2bP}\right)} + \frac{2c}{3}P^{3/2} + \frac{d}{2}P^2.
\end{equation}
Equation (\ref{eq: approx_cork}) is stated for completeness and is not used in the calculation of the fugacity coefficient.

As H$_2$, CO, and CH$_4$ approximately obey the principle of corresponding states \citep{sf87}, \cite{hp91} expressed $a$, $b$, $c$ and $d$ in equation \eqref{eq: approx_cork} in the following forms,
\begin{equation}
    \begin{split}
        a &= a_0 \frac{T_c^{5/2}}{P_c} + a_1 \frac{T_c^{3/2} T}{P_c}, \\
        b &= b_0 \frac{T_c}{P_c}, \\
        c &= c_0 \frac{T_c}{P_c^{3/2}} + c_1 \frac{T}{P_c^{3/2}}, \\
        d &= d_0 \frac{T_c}{P_c^2} + d_1 \frac{T}{P_c^2}, \\
    \end{split}
\label{eq:abcd_coeffs}
\end{equation}
where $T_c$ and $P_c$ are the critical temperature and pressure, respectively (Table \ref{tab:cp}, which is reproduced from Table 3 of \citealt{hp91}).  Table \ref{tab:other_gas} (which is reproduced from Table 2 of \citealt{hp91}) lists the empirical coefficients for computing $a$, $b$, $c$ and $d$.

\begin{table}
\begin{center}
\caption{Critical temperatures and pressures of H$_2$, CO and CH$_4$}
\label{tab:cp}
\begin{tabular}{lrr}
\hline
\hline
  gas  & $T_c$ (K) & $P_c$ (kbar)    \\ 
\hline
CH$_4$ & 190.6 & 0.0460 \\
H$_2$  & 41.2 & 0.0211  \\
CO     & 132.9 & 0.0350 \\
\hline
\hline
\end{tabular}
\end{center}
\end{table}

\begin{table}
\begin{center}
\caption{Empirical coefficients for simplified CORK EoS of H$_2$, CO and CH$_4$}
\label{tab:other_gas}
\begin{tabular}{lcr}
\hline
\hline
  $a$  & $a_0$ & $5.45963\times10^{-5}$  \\ 
       & $a_1$ & $-8.63920\times10^{-6}$  \\
  $b$  & $b_0$ & $9.18301\times10^{-4}$  \\ 
  $c$  & $c_0$ & $-3.30558\times10^{-5}$  \\
       & $c_1$ & $2.30524\times10^{-6}$  \\
  $d$  & $d_0$ & $6.93054\times10^{-7}$  \\
       & $d_1$ & $-8.38293\times10^{-8}$  \\
\hline
\hline
\end{tabular}
\end{center}
Note: $a$ and $b$ have the same units as in Tables \ref{tab:h2o} and \ref{tab:co2}, but $c$ has units of kJ kbar$^{-3/2}$ mol$^{-1}$, and $d$ has units of kJ kbar$^{-2}$ mol$^{-1}$.  To utilize these coefficients in equation (\ref{eq:abcd_coeffs}), one enters $T_c$ and $T$ in units of K and $P_c$ in units of kbar.  The coefficients $a_j$, $b_j$, $c_j$ and $d_j$ then take on the appropriate physical units.
\vspace{0.1in}
\end{table}

\section{Approximate CHOSSi system (ideal gas, ideal mixing, ignore solubility laws)}
\label{append:CHOSSi}

If we restrict ourselves to the first three net chemical reactions in equation (\ref{eq:CHOS}), assume $\gamma_i = \phi_i = 1$, ignore solubilities of gases in melt and ignore $P_{\rm O_2}$ for Dalton's law, then it is possible to obtain an analytical solution for $P$ for a CHO chemical system.  Using the first three expressions in equation (\ref{eq:partial_F}) and the equation for the carbon-to-hydrogen (C/H) ratio, one derives an explicit expression for the total surface pressure in terms of the hydrogen partial pressure,
\begin{equation}
P = P_{\rm H_2} \left( 1 + F_2 \right) \left[ 1 + \frac{2x \left( 1 + F_1 + F_4 P_{\rm H_2}^2 \right)}{1 + F_1 + F_4 P_{\rm H_2}^2 \left( 1 - 4x \right)} \right] \approx P_{\rm H_2} \left( 1 + F_2 \right),
\label{eq:P_master}
\end{equation}
where we have written $x \equiv \mbox{C/H}$ for compactness of notation.  To lowest order, the surface pressure is independent of the elemental abundance of carbon (C/H) if $x \ll 1$.

The computational recipe is as follows.
\begin{enumerate}

\item Assume a value for the partial pressure of molecular hydrogen ($P_{\rm H_2}$).  The other parameters of the system are the melt temperature $T$, the oxygen fugacity $f_{\rm O_2}$, the sulfur fugacity $f_{\rm S_2}$, C/H and the silicon-to-oxygen (Si/O) ratio.

\item Use equation (\ref{eq:P_master}) to calculate the total atmospheric surface pressure ($P$).

\item Calculate the partial pressure of carbon monoxide ($P_{\rm CO}$) using
\begin{equation}
P_{\rm CO} = \frac{P - P_{\rm H_2} \left( 1 + F_2 \right)}{1 + F_1 + F_4 P_{\rm H_2}^2}.
\label{eq:P_CO}
\end{equation}

\item Calculate the partial pressure of carbon dioxide ($P_{\rm CO_2}$), given $P_{\rm CO}$, by using the first expression in equation (\ref{eq:partial_F}).

\item Calculate the partial pressure of water ($P_{\rm H_2O}$), given $P_{\rm H_2}$, by using the second expression in equation (\ref{eq:partial_F}).

\item Calculate the partial pressure of methane ($P_{\rm CH_4}$), given $P_{\rm CO}$ and $P_{\rm H_2}$, by using the third expression in equation (\ref{eq:partial_F}).

\item Calculate the partial pressure sulfur dioxide ($P_{\rm SO_2}$) using equation (\ref{eq:SO2}).

\item Post-process for the partial pressure of hydrogen sulfide ($P_{\rm H_2S}$), given $P_{\rm H_2O}$, by using the fourth expression in equation (\ref{eq:partial_F}), even though it was not formally part of the system of equations.

\item Using the equation for the silicon-to-oxygen ratio (Si/O), post-process for the partial pressure of silicon monoxide using the following expression,
\begin{equation}
P_{\rm SiO} = \frac{y \left( P_{\rm CO} + P_{\rm H_2O} + 2 P_{\rm CO_2} + 2 P_{\rm SO_2} + 2 P_{\rm O_2} \right)}{1 - y + F_9 P_{\rm H_2}^2},
\end{equation}
where we have written $y \equiv \mbox{Si/O}$ and $F_9 \equiv F_8/F_2$ for compactness of notation, even though it was not formally part of the system of equations.

\item Post-process for the partial pressure of silane ($P_{\rm SiH_4}$), given $P_{\rm SiO}$, $P_{\rm H_2}$ and $P_{\rm H_2O}$, by using the fifth expression in equation (\ref{eq:partial_F}), even though it was not formally part of the system of equations.

\end{enumerate}

For secondary atmospheres, the preceding computational recipe is identical except for the first two steps: instead of assuming $P_{\rm H_2}$, one assumes $P$.  One then has to solve the following cubic equation for $P_{\rm H_2}$ numerically,
\begin{equation}
P_{\rm H_2}^3 F_4 \left( 1 + F_2 \right) \left( 1 - 2x \right) - P_{\rm H_2}^2 F_4 \left( 1 - 4x \right) P + P_{\rm H_2} \left(1 + F_1 \right) \left( 1 + F_2 \right) \left( 1 + 2x \right) - \left(1 + F_1 \right) P = 0,
\label{eq:secondary}
\end{equation}
where it is apparent that the partial pressure of molecular hydrogen is, to lowest order, independent of $x$ if $x \ll 1$.

While this CHOSSi chemical system is not sophisticated enough for generating results, it provides adequate first guesses for more advanced chemical systems that require numerical iteration to converge to the solutions.  In practice, we find that the post-processed partial pressures for hydrogen sulfide, silicon monoxide and silane are not good first guesses when the volume mixing ratio of water is non-negligible, and using the partial pressure of water instead serves as good first guesses for these three species.

\section{Iron-w\"{u}stite (IW) Buffer}
\label{append:IW}

\begin{figure}
\begin{center}
\vspace{-0.2in} 
\includegraphics[width=0.6\columnwidth]{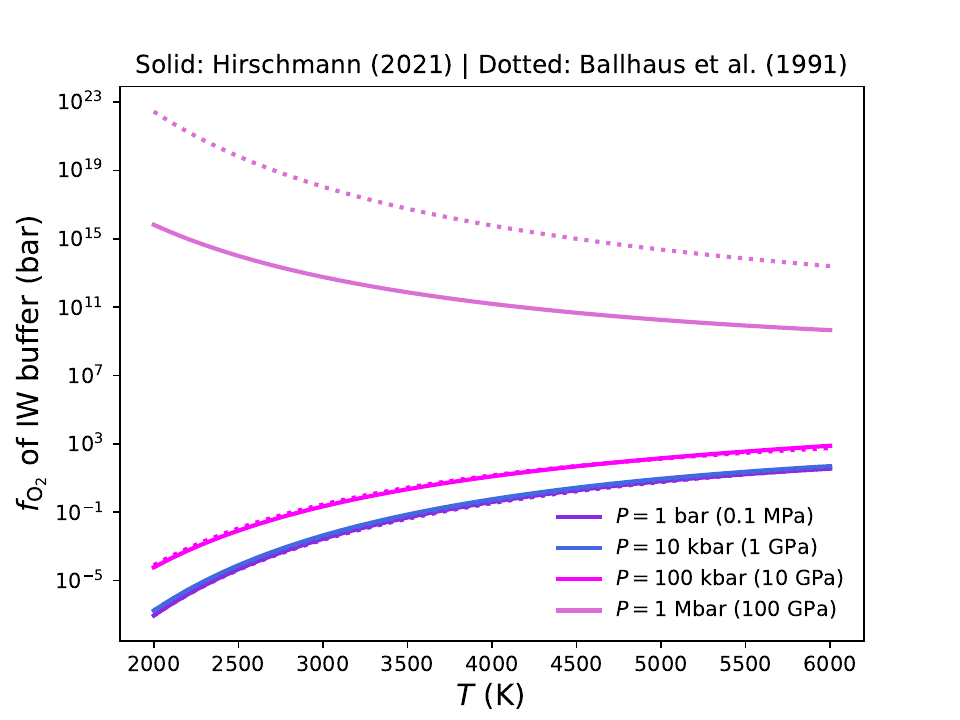}
\end{center}
\vspace{-0.1in}
\caption{The oxygen fugacity (in bar) of the IW buffer using the empirical formulae of \cite{ballhaus91} and \cite{hirschmann21}, which are shown as dotted and solid curves, respectively.}
\vspace{0.2in}
\label{fig:IW}
\end{figure}

The empirical formula of \cite{ballhaus91} is based on the experimental data of \cite{oneill87}, which is valid for $833 \le T \le 1450$ K.  \cite{ballhaus91} used these data to produce an empirical fitting formula for IW (see their Table 2), which is restated in equation (27) of \cite{th24}.  \cite{ballhaus91} applied theoretical pressure corrections to derive this formula and its pressure range of validity is unclear. 

\cite{hirschmann21} extended the IW empirical formula to $1000 \le T \le 3000$ K and pressures of $100 \mbox{ kPa} \le P \le 100 \mbox{ GPa}$.  However, we have already noted that this does not prevent the oxygen fugacity from exceeding the total pressure under certain conditions. 

Figure \ref{fig:IW} shows the oxygen fugacity in absolute units (bar) of the IW buffer.  The pressure dependence of IW is negligible between 1 bar (Earth-like surface pressures) and 10 kbar (sub-Neptune-like surface pressures).  At pressures relevant for sub-Neptunes ($\lesssim 100$ kbar), the formulae of \cite{ballhaus91} and \cite{hirschmann21} agree well (Figure \ref{fig:IW}).  Unsurprisingly, these formulae diverge at $100 \mbox{ GPa} = 1$ Mbar, but such a pressure is irrelevant for sub-Neptunes.  At this pressure, the oxygen fugacity \textit{decreases} with temperature according to the empirical formulae.

\label{lastpage}

\end{document}